\documentclass[%
 aip,
 amsmath,amssymb,
 reprint,%
]{revtex4-1}

\usepackage{graphicx}% Include figure files
\usepackage{dcolumn}% Align table columns on decimal point
\usepackage{bm}% bold math
\usepackage{color,soul}
\usepackage[utf8]{inputenc}
\usepackage[T1]{fontenc}
\usepackage{mathptmx}

\begin{document}

\preprint{AIP/123-QED}

\title{Superconducting quantum many-body circuits for quantum simulation and computing}

\author{Samuel A. Wilkinson}
\affiliation{Friedrich-Alexander University Erlangen-N\"urnberg (FAU), Department of Physics, Erlangen, Germany}
\author{Michael J. Hartmann}
\affiliation{Friedrich-Alexander University Erlangen-N\"urnberg (FAU), Department of Physics, Erlangen, Germany}
\affiliation{Max Planck Institute for the Science of Light, Erlangen, Germany}

\date{\today}% It is always \today, today,

\begin{abstract}
Quantum simulators are attractive as a means to study many-body quantum systems that are not amenable to classical numerical treatment. A versatile framework for quantum simulation is offered by superconducting circuits. In this perspective, we discuss how superconducting circuits allow the engineering of a wide variety of interactions, which in turn allows the simulation of a wide variety of model Hamiltonians. In particular we focus on strong photon-photon interactions mediated by nonlinear elements. This includes on-site, nearest-neighbour and four-body interactions in lattice models, allowing the implementation of extended Bose-Hubbard models and the toric code. We discuss not only the present state in analogue quantum simulation, but also future perspectives of superconducting quantum simulation that open up when concatenating quantum gates in emerging quantum computing platforms. 
\end{abstract}

\maketitle

\section*{Introduction}

The dimension of the Hilbert space of a quantum system grows exponentially in the number of constituents (e.g. particles, lattice sites) of that system, so that explicitly computing the eigenspectrum or time evolution of a quantum system of even quite modest size quickly becomes intractable on a classical computer. For example, the Hilbert space of a lattice of $N$ spins with spin-1/2 scales as $2^N$, so that while calculations of three or four spins are easy to perform classically, a lattice of 50 spins naively requires over 140 TB of data just to store the state of the system. ~50 spins represents the upper limit of what is currently possible on a classical computer; a system of just 60 spins currently presents an insurmountable task. While there are many useful approximation techniques \cite{Kuramoto2020} one can employ to alleviate this problem, in general computing useful quantities for quantum many-body systems on a classical computer is not possible unless some major simplifying assumptions can be made. In particular, no general-use numerical tools exist for analysing a large many-body quantum system which is highly-entangled or driven far-from-equilibrium.

In 1982, Richard Feynman suggested this problem may be solved by using controllable quantum systems to simulate the dynamics of other quantum systems \cite{Feynman1982}. In this way, we use the exponential growth of Hilbert space to our advantage -- as the amount of information needed to describe the system increases exponentially, so too must the amount of information which the system can describe. This suggestion, along with other preliminary works \cite{Lloyd1996}, initiated the research program known as \textit{quantum simulation}\footnote{Feynman's suggestion is also the basis of the broader research paradigm of quantum computing. The topic of quantum computing in general is extremely extensive, and an excellent introduction is given in the classic textbook \cite{Nielsen2010}}.

There are two main flavours of quantum simulation. \textit{Analogue} quantum simulation (AQS) involves the construction of a controllable, engineered quantum system which mimics the physical behaviour of the system to be simulated. More precisely, performing an analogue quantum simulation of a system $H$ means operating a device in such a way that it's dynamics are governed by an effective Hamiltonian which approximates $H$.
\textit{Digital} quantum simulation (DQS) \cite{Lloyd1996}, on the other hand, involves designing a sequence of unitary gates to approximate unitary evolution under the Hamiltonain being simulated. This is generally done using the Suzuki-Trotter formula through a process called ``Trotterization". Both approaches have also been combined into a strategy coined "digital-analogue" quantum simulation, where evolution steps of analogue quantum simulations are concatenated in a Trotter sequence \cite{Langford2017,Lamata2018}.

In the review parts of this article, we will focus mainly on AQS. However, we shall discuss DQS in more detail when we describe the future of quantum simulation -- as we expect this concept to become increasingly important due to rapid developments in building digital quantum computers.

\section*{Superconducting Circuits}

\begin{figure*}[tb!]
	\includegraphics[width=\textwidth]{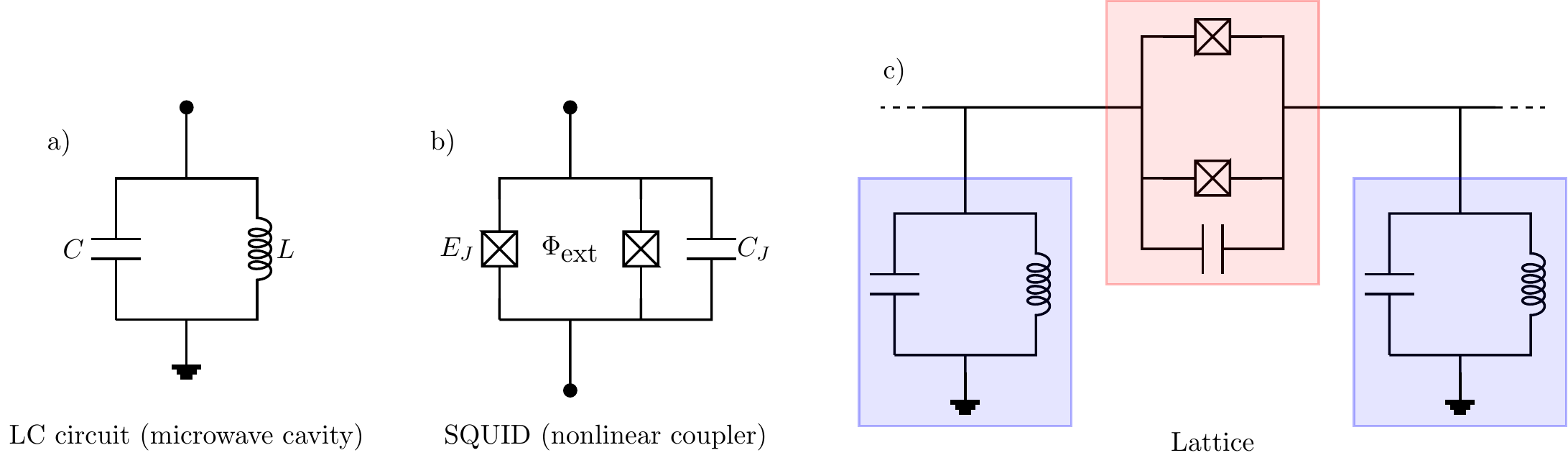}
	
	\caption{\label{Fig:circuits} Circuit diagrams for a) an LC-oscillator, b) a capacitively-shunted SQUID and c) a lattice of LC-oscillators connected via SQUIDs. The LC-oscillators (blue) act as microwave resonators, while the SQUIDs (red) provide a nonlinear coupling. }
\end{figure*}

Why are superconducting circuits such a successful platform for quantum simulation?
Detailed information of the state or properties of the simulator must be available to the scientist running the simulation (otherwise the simulator would offer little advantage over performing experiments on the system itself). The simulator must furthermore be scalable to sizes exceeding what can be classically simulated, and must exhibit long coherence times -- at least of the order of the time needed to enact the simulation. These criteria are well met by superconducting qubits \cite{Wendin2017,Martinis2020,Carusotto2019,Blais2020}, which have coherence times on the order of 10-100$\mu$s and can be individually measured via dispersive shift \cite{Kjaergaard2019,Krantz2019}, allowing readout at the level of individual lattice sites.

We shall primarily consider the framework of circuit quantum electrodynamics (CQED), in which superconducting LC circuits act as resonators, hosting microwave photons, as shown in Fig~\ref{Fig:circuits} a).
Photon-photon interactions can be mediated by nonlinear elements, such as superconducting qubits or SQUIDs, shown in Fig.~\ref{Fig:circuits} b).
This is analogous to the situation in cavity QED, where atoms mediate interactions between optical photons.
Circuit QED has an advantage over its cavity counterpart because in a superconducting circuit the photons have a wavelength on the order of 10 mm, so that building many resonators with a size comparable to the wavelength is well within the capabilities of modern fabrication techniques.
Thus the construction of large, coherently coupled arrays with high finesse is much easier than in the optical case.

While there are many different superconducting qubit designs, the most relevant in the context of near-term quantum simulation is the \textit{transmon} \cite{Koch2007}. This consists of a single Josephson junction (or a SQUID loop) shunted by a large capacitance. The junction provides non-linearity, while the shunt capacitance flattens the charge dispersion and thus protects the qubit from charge noise. Transmons are favoured because of their long coherence times \cite{Houck2009} and the flexibility of their couplings (as we shall see below). Excitations of a transmon are microwave photons (quantized current-voltage oscillations across the junction), with non-linearity determined by the ratio $E_C/E_J$ between the Josephson energy $E_J$ associated with the coherent tunnelling of Cooper pairs across the junction, and the charging energy $E_C = (2e)^2/2C$ where $C$ is the total capacitance (sum of shunt and junction capacitances). To be well-protected from charge noise, the charging energy must be small and consequently the non-linearity is small -- however, as we shall see, the non-linearity is still large enough to produce strong interactions and highly non-trivial physics\footnote{An alternative qubit design, the capacitively-shunted flux qubit \cite{Mooij1999,Yan2016}, has much larger non-linearity with similar single-qubit coherence times to transmons. However, presently these flux qubits cannot be coupled with the scalability and flexibility of transmons without introducing additional noise and thus reducing coherence.} .

A key feature of superconducting circuits is their flexibility. Superconducting cavities, qubits and transmission lines may be coupled with DC SQUIDS, which allows a wide variety of interactions to be tuned continuously or switched on-and-off via constant or oscillating external magnetic fluxes \cite{Neumeier2013}.
Below we shall examine in detail some of the couplings that can be implemented (such as on-site, nearest-neighbour and four-body interactions), and some of the model Hamiltonians this allows us to simulate (such as the Bose-Hubbard model and the toric code).
In all of the cases discussed below, the crucial ingredient is a nonlinear coupling element.

\section*{Engineering Interactions and Hamiltonians}

Whereas linear couplings between circuits, which enable propagation of excitations, can be realized with linear inductors or capacitances, the simplest way to implement controllable, nonlinear coupling between photon modes is to connect two circuit nodes via a Josephson junction \cite{Jin2013, Neumeier2013,Collodo2019}.
This approach is easily scalable to large lattices, where the nodes of the lattice can be resonators or qubits and the edges are Josephson junctions \footnote{With one caveat -- one must be careful not to create closed superconducting loops, which lead to flux quantization and make the system more sensitive to magnetic flux noise. This can be avoided by introducing a capacitor into the edges, such that the loop is broken.}.
Despite its simplicity, this method is already sufficient to implement linear and correlated hopping between cavities, as well as Kerr and cross-Kerr non-linearities \cite{Collodo2019} (equivalently, on-site and nearest-neighbour interactions in the lattice) in a way that these couplings can be tuned independently.
This allows the engineering of the Bose-Hubbard model \cite{Fisher1989,Sachdev2011}, as well as extended Bose-Hubbard models which exhibit a richer phase diagram \cite{Rossini2012}.

\begin{figure}[tb!]
	\includegraphics[width=\columnwidth]{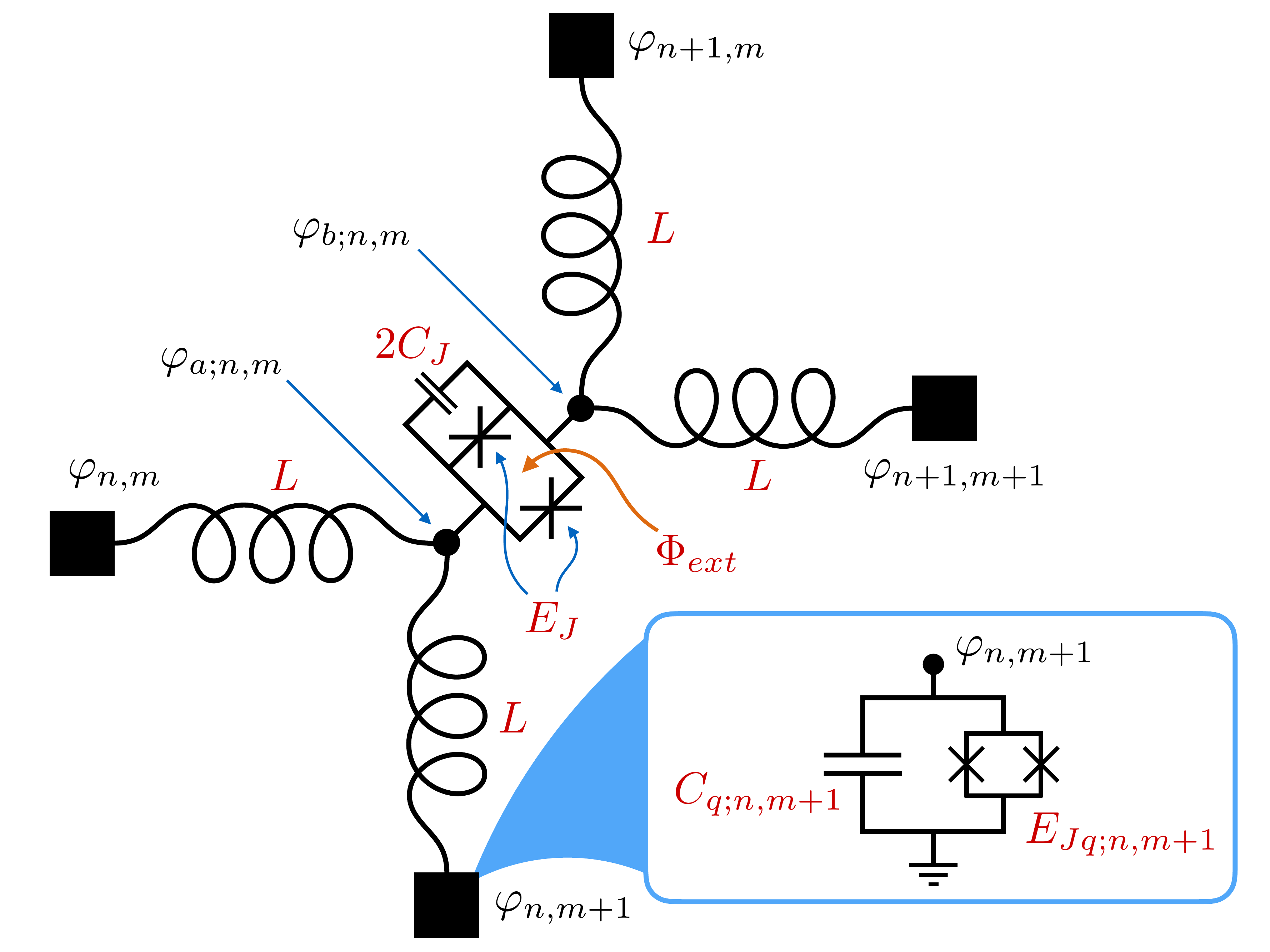}
	\caption{\label{Fig:TocirCircuit} Circuit of a physical cell for implementing a Toric Code Hamiltonian. The rectangles at the corners represent qubits with different transition frequencies, see inset for the circuit of each qubit. The qubits form a two dimensional lattice and are thus labelled by indices $(n,m)$ indicating their position along both lattice axes. Qubits $(n,m)$ and $(n,m+1)$ [$(n+1,m)$ and $(n+1,m+1)$] are connected via the inductances $L$ to the node $(a;n,m)$ [$(b;n,m)$]. The nodes $(a;n,m)$ and $(b;n,m)$ in turn are connected via a dc-SQUID with an external magnetic flux $\Phi_{ext}$ threaded through its loop. The Josephson junctions in the dc-SQUID have Josephson energies $E_{J}$ and capacitances $C_{J}$, which include shunt capacitances. Figure adapted from \cite{Sameti2017}}.
\end{figure}

Superconducting circuits even allow engineering of interactions that do not occur in nature (at least at these energy scales). Among those interactions are many-body interactions as they appear in toric code, which is the prototypical model for fault-tolerant quantum computing \cite{Kitaev2002,Kitaev2009,Fowler2012} and an example of a Z$_2$ lattice gauge theory. The challenge in engineering the toric code is that it requires implementing four-body interactions while suppressing two-body interactions. 
In superconducting circuits this can be done by coupling qubits with a dc SQUID and driving the SQUID with an oscillating magnetic flux \cite{Sameti2017}, see Fig. \ref{Fig:TocirCircuit} for the circuit.

The four qubits participating in the interaction all have different transition frequencies, $\omega_j$, $j\in \{1,2,3,4\}$.
By modulating the flux through the SQUID at a frequency $\omega_d = \omega_1 + \omega_2 + \omega_3 + \omega_4$, we drive the four-body interaction $a_1^\dagger a_2^\dagger a_3^\dagger a_4^\dagger$, where $a_j^\dagger$ creates an excitation on the $j^\textrm{th}$ qubit. More generally, for any four-body interaction $\prod_{j=1}^4 b_j$, $b_j \in \{a_j,a_j^\dagger\}$ can be implemented by some choice of drive frequency $\omega_d = \sum_{j=1}^4 s_j\omega_j$, $s_j = \pm 1$.
Thus, by carefully selecting the driving frequency we are able to implement arbitrary four-body interactions, while unwanted two-body interactions arising from the drive can be made negligible by choosing the differences between transition frequencies to be much larger than the relevant coupling strength.
Periodic driving of a coupling element at frequencies comparable to qubit transition frequencies has already been demonstrated \cite{Niskanen2007,Roushan2017}, and realising the minimal toric code on 8 qubits in this setup is within current experimental capabilities. Moreover, periodic driving of circuit element has already been applied in experiments to simulate coupling strengths that would otherwise be inaccessible in the device, such as so called ultra-strong coupling where excitation numbers are no longer conserved \cite{Braumuller:2017zl}.

The idea of periodically driving a system in order to qualitatively alter its long-time dynamics is formalized and generalized in the framework of \textit{Floquet engineering} \cite{Oka2019}.
While the CQED architecture allows the construction of near-arbitrary lattices, applying techniques of Floquet engineering in this context allows for the implementation of arbitrary spin-spin interactions \cite{Sameti2019}.
As a demonstration of the power of this method, it was shown that Floquet engineering in superconducting circuits can simulate the Kitaev honeycomb model, and that spin-spin couplings along different axes can be tuned interdependently so that both the Abelian and the non-Abelian phase can be explored \cite{Sameti2019}.

Fine control over not only the magnitude, but also the complex phase of photon hopping offers the possibility of engineering synthetic magnetic fields.
When simulating quantum lattice models with a magnetic field on a classical computer, one typically uses a technique known as Peierl's substitution \cite{Peierls1933}, whereby one adds a complex phase to the hopping terms such that the sum of phases around a plaquette of the lattice is equal to the magnetic flux through that plaquette.
In AQS, we can use this same technique by physically controlling the photon tunnelling phase, causing photons to behave like charged particles in a magnetic field \cite{Fang2012,Roushan2017}.
Alternatively, magnetic fields can be synthesized by breaking time-reversal symmetry, which can be achieved by on-chip circulators \cite{Koch2010,Muller2018}.
Exactly analogous to the topological phases that have been observed in solid state systems under magnetic fields, synthetic magnetic fields in superconducting circuits allow the engineering of topological photonics \cite{Ozawa2019}, so that topologically non-trivial materials can be simulated by a photonic device.

Superconducting circuits can thus be used to engineer a large variety of effective Hamiltonians. Yet, they differ strongly from other platforms such as ultra-cold atoms \cite{Bloch2008} in the way they are affected by dissipation. This aspect need not only be an imperfection, but also gives rise to very intriguing scenarios as we discuss next.

\section*{Driven-dissipative Regimes}

\begin{figure}[tb!]
	\includegraphics[width=.8\columnwidth]{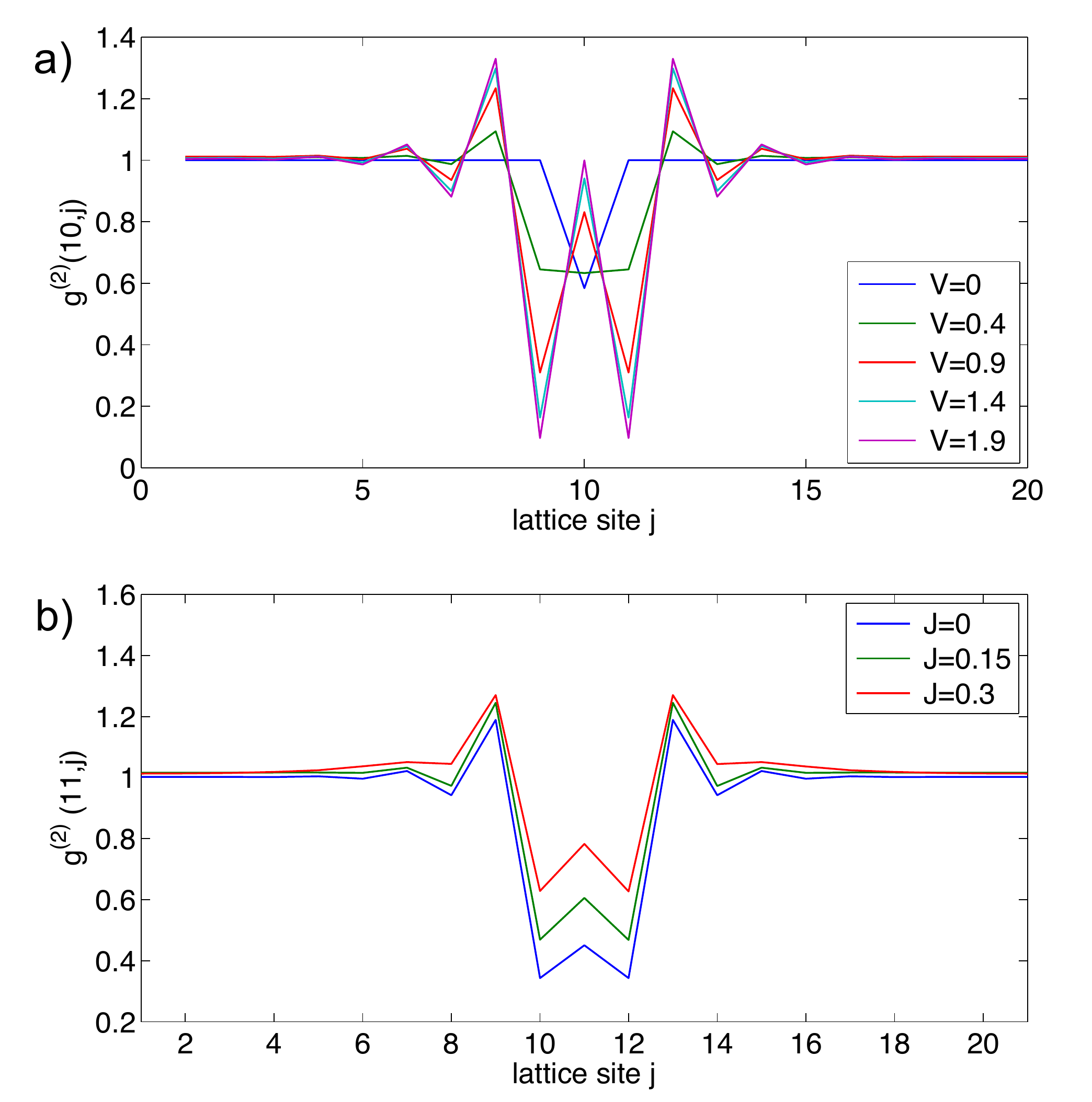}
	\caption{\label{Fig:corssg2} Spatially resolved density-density correlations between two lattice sites $j$ and $l$, $g^{(2)}(j,l)$. a) Results of a Matrix Product State simulation for vanishing tunneling rate between resonators with onsite interactions $U=0.4$, drive amplitude $\Omega = 0.4$, and cross Kerr interactions $V$ as in the legend (the dissipation rate is set to 1) , b) as in plot a, but with $U=V=1$ and varying tunneling rate $J$ as given in the legend. Figure adapted from \cite{Jin2013}}.
\end{figure}

As the elementary excitations in superconducting circuits are photons, quantum simulators using this technology are inevitably affected of photon losses.
Photons decay out of their cavities, and need to be continually re-injected by an external drive.
This means that superconducting circuit lattices are naturally studied from the perspective of non-equilibrium driven-dissipative systems.
For interactions to be significant their coupling needs to be stronger than the photon decay rates, and equilibration is difficult, if not impossible to achieve.
While this may seem a hindrance at first, it has been turned into a strength of photonic approaches to AQS, as it opens up the possibility of exploring the realm of driven-dissipative dynamics and allows us to study quantum phases and phase transitions that are unique to far-from-equilibrium systems \cite{Eisert2015}.
Dissipative phase transitions were initially observed in two coupled microwave resonators \cite{Raftery2014}, and more recently in a chain of 72 superconducting cavities coupled via transmons \cite{Fitzpatrick2017}.
However, such far-from-equilibrium systems have only recently begun to be understood, and present many challenges both theoretically and practically.
Given that the purpose of quantum simulation is to study systems which are difficult to simulate classically or about which we have limited information experimentally, far-from-equilibrium systems are excellent targets for AQS.

An important early step in establishing this was to demonstrate that photon crystallization can occur even in a driven-dissipative system \cite{Hartmann2010}.
A more detailed analysis revealed that the phase diagram for a coupled cavity array can be incredibly rich, even for the simple case where cavities are coupled by static transmons, where uniquely nonequilibrium phases appear in between photon crystallization and delocalization \cite{Jin2013}.

\begin{figure}[tb!]
	\includegraphics[width=\columnwidth]{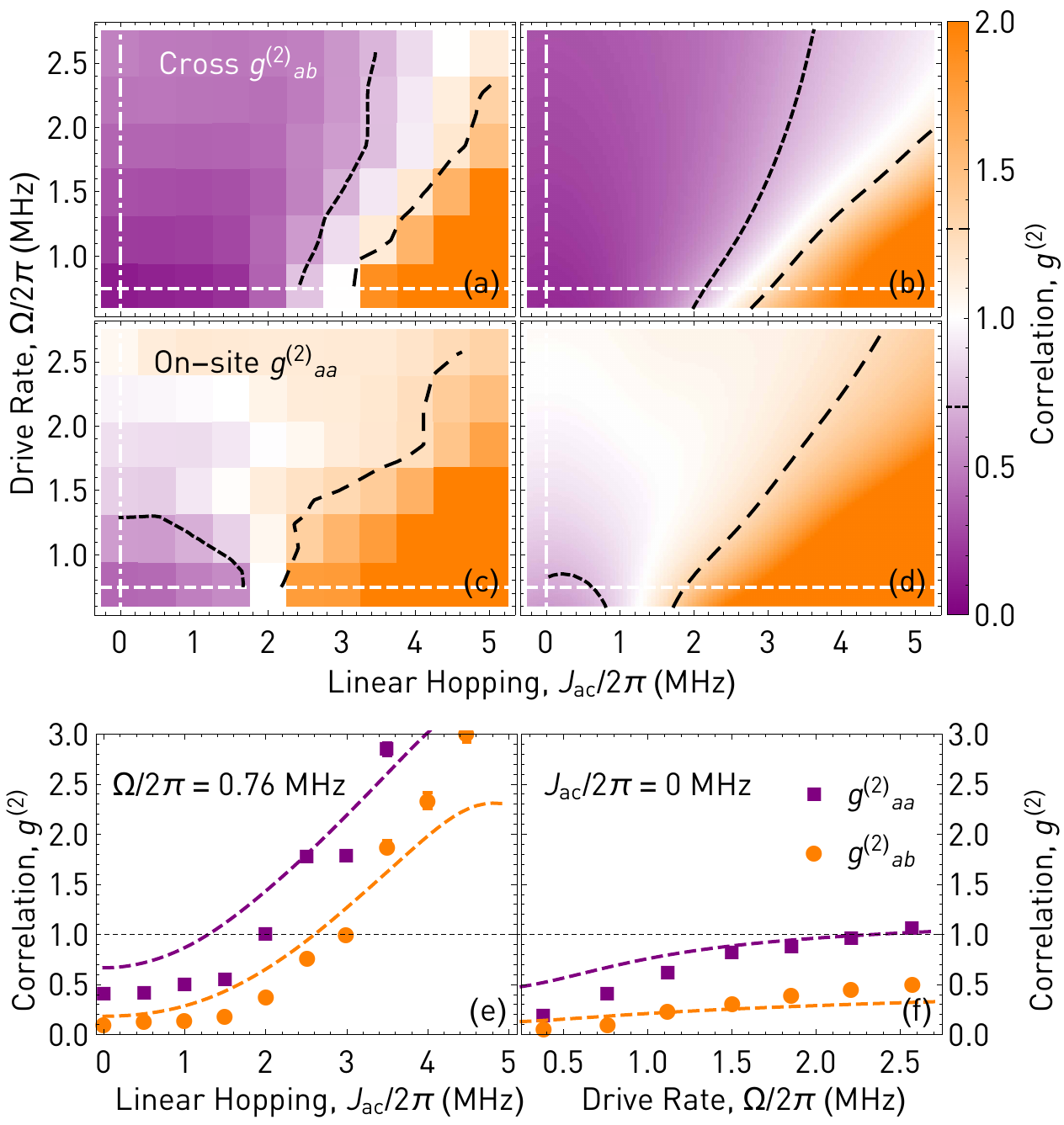}
	\caption{\label{Fig:corssg2_exp} On-site and cross $g^{(2)}$-functions in a setup of two coupled resonators. (a) and (c) are measured $g^{(2)}$-functions as functions of linear hopping rate $J_{ac}$ and drive amplitude $\Omega$. (b) and (d) are the corresponding results from numerical simulations.  (e) and (f) are cuts at constant  $J_{ac}$ and $\Omega$. Figure adapted from \cite{Collodo2019}.}
\end{figure}

We can even design lattices where nearest-neighbour repulsion is stronger than on-site repulsion (i.e., cross-Kerr non-linearity exceeds on-site Kerr non-linearity) \cite{Jin2013,Kounalakis2018}.  This gives rise to a phase which displays both global coherent phase oscillations and checkerboard ordering of photon density, implying a nonequilibrium supersolid \cite{Jin2013}, see Fig. \ref{Fig:corssg2}. Proof-of-principle experiments have been conducted in two-site systems; in \cite{Collodo2019}, a circuit of two superconducting resonators coupled by a SQUID, which acts as a nonlinear coupler mediating on-site and cross-Kerr interactions, as well as linear hopping between the two cavities which can be tuned \textit{in situ} via static and oscillating external magnetic fields.
In that experiment a crossover between a delocalized and an ordered phase was observed as the linear hopping rate was tuned, see Fig. \ref{Fig:corssg2_exp}.
At small hopping and low drive, the system exhibits anti-bunched photon statistics, which is interpreted as a finite-system equivalent of spontaneous symmetry breaking and photon crystalization.
This suggests that phase transitions of light will be observable in larger superconducting circuit lattices.

The versatility of superconducting circuits allows the engineering not only of lattice systems, but also continuous, spatially-extended systems, such as nonlinear waveguides.
In CQED, waveguides commonly take the form of superconducting coplanar waveguide resonators (CWR).
By taking a CWR and interrupting it periodically with Josephson junctions, we can engineer strong photon-photon interactions \cite{Leib2014}.
When a single mode is driven with strength $\Omega$, this systems displays novel synchronization behaviour, where all modes are suppressed at low drive $|\Omega| < \Omega^*$, but there is a sudden synchronized switch-on at some threshold drive $\Omega^*$ that leads to large occupancies of all modes at once.

An issue arising from the driven-dissipative nature of photonic systems is the difficulty of engineering a chemical potential for photons. Photon number is typically not a conserved quantity in experiments, and furthermore photon trapping times may be similar to or even significantly less than equilibration time. Consequently, there is no well-defined notion of a ``chemical potential" in photonic systems. This remains an open from for all forms of AQS based on photons \cite{Hartmann2016}.

Nonetheless, the driven-dissipative nature of superconducting circuits does not rule out the possibility of using such devices as AQSs of equilibrium phases of matter. In a non-equilibrium setting, dissipation can be used to stabilize phases originally observed in equilibrium systems.
For example, a recent experiment has shown a Mott insulating phase in eight coupled transmon qubits that was stabilized by dissipation \cite{Ma}.

Having discussed some of the major achievements of superconducting quantum simulation to date, let us now look to what we can expect from the future, and what further avenues of research might prove most fruitful.

Previous experiments in AQS with superconducting circuits have focused on devices with very few lattice sites, or on one-dimensional chains.
However, these are systems which can be studied using classical numerical techniques, and can only serve as proof-of-concept experiments.
To truly unleash the power of quantum simulation, we need to expand our efforts to larger systems, including two- (and possible three-) dimensional lattices.
This is difficult, partly because experiments are restricted to planar circuits, due to the physical limitations of on-chip design.
However, elements such as air-bridges \cite{Chen2014a,Dunsworth2018} or layered devices can allow us to overcome this restriction and implement non-planar designs. 
Periodic boundary conditions for an effectively 3D array have been realised in linear superconducting circuits \cite{Ningyuan2015,Lu2019}, allowing the simulation of bulk systems in a finite device.
Currently lattice design in superconducting circuits is quite flexible, but the implementation of non-planar circuits promises to allow the construction of near-arbitrary lattices.

In fact, the flexibility of superconducting circuits means that it should be possible not only to simulate arbitrary two-dimensional \textit{Euclidean} geometries, but also curved spaces.
This is because, although the circuit is confined to a Euclidean plane, the simulated physics depends only on the geometry implied by the circuit topology and coupling strengths.
If a lattice in a curved geometry can be projected onto a Euclidean plane, then that curved geometry can be simulated in a planar circuit.
Proof-of-concept experiments have already been performed on small hyperbolic lattices \cite{Kollar2019}.

As with all applications involving superconducting qubits, quantum simulators are limited by the quality of the qubits themselves. While circuit fabrication technology has improved greatly over the years, complications such as two level system (TLS) defects in the junctions and unwanted variation in the junction parameters unavoidably introduce decoherence and disorder respectively. As our understanding of the microscopic details of the fabrication process becomes more advanced, it may become possible to remove or at least reduce these problems \cite{Cyster2020}.

Unlike standard gate-based quantum computations, in AQS there is no natural notion of error-correction.
We expect certain simulations to be more robust to error than others -- for example, in the simulation of driven-dissipative system, photon loss is not an error, but rather a feature of the simulation.
Methods have been developed to diagnose the robustness of AQS, and found that symmetries can be exploited to increase the reliability of simulations \cite{Sarovar2017}.
However, no realistic device will be free from imperfections, and thus to ensure that an AQS performs reliably, we must typically resort to benchmarking protocols \cite{Heyl2019,Derbyshire2020}.

\section*{Digital Quantum Simulation}

If the ultimate goal is to produce a product of broad utility, then we had best consider the flexibility of our designs. 
While AQSs are highly tunable and controllable, in general we expect each circuit to be capable of simulating only a single Hamiltonian or a relatively small class of Hamiltonians (albeit in many different regions of parameter space).
To study a different model system, an entirely new circuit would have to be fabricated.
DQS, on the other hand, can be made to be programmable, such that a single device can simulate any system. This approach has now become increasingly feasible as the execution of circuits with significant numbers of quantum gates on more than 50 qubits have been demonstrated \cite{Arute2019}. 

Important milestones in DQS on superconducting circuits have included showing experimentally that a superconducting circuit can be used to simulate lattices of spins \cite{Salathe2015,Heras2014} and fermions\cite{Barends2015,Heras2015} -- both important models in condensed matter physics and both of potential interest to industry. Recent experiments on superconducting circuits have also demonstrated digital quantum simulation of the quantum Rabi and Dicke models \cite{Mezzacapo2014,Langford2017}. However, current DQS devices still fall short of the accuracy and scalability that will be necessary for DQS to outcompete classical computational methods \cite{McClean2016}.

DQS also benefits from sharing an architecture with universal quantum computation.
Thus improvements in quantum computing technology with the aim of, say, implementing Shor's factoring algorithm \cite{Shor2002} or efficiently performing linear algebra subroutines \cite{Harrow2009}, will also benefit DQS.
General efforts towards scalable, fault-tolerant quantum computing will improve DQS almost as an afterthought.

Additionally, DQS is inherently less platform-dependent than analogue simulation, and therefore there is nothing which fundamentally prevents protocols developed for superconducting qubits from being applied on other quantum simulation architectures such as e.g. trapped ions \cite{Blatt2012}.

However, at least at present, the accuracy of DQS is limited by so-called ``Trotter errors" \cite{Heyl2019}.
To simulate evolution under a Hamiltonian containing many terms, DQS makes use of the Trotter-Suzuki formula to approximately separate the total evolution into a sequence of evolutions under each term independently.
This allows the evolution to be implemented as a series of quantum gates, at the cost of some error -- the Trotter error -- arising from the Trotter-Suzuki formula \cite{Tranter2019,Childs2019} .
The Trotter error can be made smaller by increasing the number of independent evolutions in the sequence, but this increases the number of gates needed to perform the simulation.
On current and near-term hardware, gates are inherently noisy and each have some finite error of their own.
So that while increasing the number of gates reduces Trotter error, it compounds errors arising from imperfect gates, which quickly leads to unreliable results.

In the longer term DQS will also profit from error correction strategies that have been developed for quantum computing. While current hardware hasn't yet reached the size and gate fidelities for error correction to be a really useful approach, this possibility of DQS can become an important feature in the future, boosting the simulation capabilities significantly. Moreover DQS schemes may in the future also benefit from the integration of many-body coupler circuits as here discussed in the context of engineering specific many-body interactions for AQS, see e.g. figure \ref{Fig:TocirCircuit}. Such circuits could allow for single step multi-qubit gates, thus reducing gate count of algorithms.

DQS can not only be applied to simulating the time evolution generated by a specific Hamiltonian, but can also be extended to finding gate sequences that prepare the ground state of a target Hamiltonian in a variational quantum algorithm \cite{McClean2016}. Particularly in the commencing era of ``Noisy Intermediate Scale Quantum" (NISQ) computing \cite{Preskill2018}, such variational quantum algorithms are promising early applications of quantum computers. These are hybrid algorithms that execute a moderate size gate sequence on a quantum chip and optimize the parameters that this gate sequence depends on via a classical optimizer that runs on a classical computer. These algorithms are robust against the imperfections of current quantum computers as the required gate sequences remain relatively short but are run many times so that errors can be mitigated using statistical techniques. 

This concept has already been successfully applied in several platforms, including superconducting circuits \cite{OMalley2016,Kandala:2017fa,Kokail:2019fw}. It remains an open whether quantum variational algorithms can be extended to a size where they could make predictions that are no longer feasible with classical computing architectures.

\section*{Conclusions}

In this work we have attempted to give a sense for the flexibility of superconducting circuits, and the great amount of control we can exert over superconducting circuits. Interactions and lattice geometries can be constructed almost at will, which opens up the exciting possibility of simulating an enormous variety of many-body systems, including strongly-interacting and far-from-equilibrium systems. 

In the coming years we expect digital approaches to quantum simulations which can run on near-term quantum hardware to be the most important research direction for superconducting circuit-based devices. This endeavour goes hand in hand with applications of quantum computers in material science and quantum chemistry, which is expected to be the area where the first big impact of quantum computing will be seen. One of the most exciting scientific questions for the coming years is therefore whether digital quantum simulation will allow us to answer a scientific question that we haven't been able to answer using classical computing architectures.

\vspace{20pt}

\section*{Acknowledgements}

This work has received funding from the European Union's Horizon 2020 research and innovation programme under grant agreement No 828826 ``Quromorphic"

\bibliography{Bib}

\begin{thebibliography}{73}%
\makeatletter
\providecommand \@ifxundefined [1]{%
 \@ifx{#1\undefined}
}%
\providecommand \@ifnum [1]{%
 \ifnum #1\expandafter \@firstoftwo
 \else \expandafter \@secondoftwo
 \fi
}%
\providecommand \@ifx [1]{%
 \ifx #1\expandafter \@firstoftwo
 \else \expandafter \@secondoftwo
 \fi
}%
\providecommand \natexlab [1]{#1}%
\providecommand \enquote  [1]{``#1''}%
\providecommand \bibnamefont  [1]{#1}%
\providecommand \bibfnamefont [1]{#1}%
\providecommand \citenamefont [1]{#1}%
\providecommand \href@noop [0]{\@secondoftwo}%
\providecommand \href [0]{\begingroup \@sanitize@url \@href}%
\providecommand \@href[1]{\@@startlink{#1}\@@href}%
\providecommand \@@href[1]{\endgroup#1\@@endlink}%
\providecommand \@sanitize@url [0]{\catcode `\\12\catcode `\$12\catcode
  `\&12\catcode `\#12\catcode `\^12\catcode `\_12\catcode `\%12\relax}%
\providecommand \@@startlink[1]{}%
\providecommand \@@endlink[0]{}%
\providecommand \url  [0]{\begingroup\@sanitize@url \@url }%
\providecommand \@url [1]{\endgroup\@href {#1}{\urlprefix }}%
\providecommand \urlprefix  [0]{URL }%
\providecommand \Eprint [0]{\href }%
\providecommand \doibase [0]{http://dx.doi.org/}%
\providecommand \selectlanguage [0]{\@gobble}%
\providecommand \bibinfo  [0]{\@secondoftwo}%
\providecommand \bibfield  [0]{\@secondoftwo}%
\providecommand \translation [1]{[#1]}%
\providecommand \BibitemOpen [0]{}%
\providecommand \bibitemStop [0]{}%
\providecommand \bibitemNoStop [0]{.\EOS\space}%
\providecommand \EOS [0]{\spacefactor3000\relax}%
\providecommand \BibitemShut  [1]{\csname bibitem#1\endcsname}%
\let\auto@bib@innerbib\@empty
%</preamble>
\bibitem [{\citenamefont {Kuramoto}(2020)}]{Kuramoto2020}%
  \BibitemOpen
  \bibfield  {author} {\bibinfo {author} {\bibfnamefont {Y.}~\bibnamefont
  {Kuramoto}},\ }\href@noop {} {\emph {\bibinfo {title} {{Quantum Many-Body
  Physics}}}}\ (\bibinfo  {publisher} {Springer Japan},\ \bibinfo {year}
  {2020})\BibitemShut {NoStop}%
\bibitem [{\citenamefont {Feynman}(1982)}]{Feynman1982}%
  \BibitemOpen
  \bibfield  {author} {\bibinfo {author} {\bibfnamefont {R.~P.}\ \bibnamefont
  {Feynman}},\ }\bibfield  {title} {\enquote {\bibinfo {title} {{Simulating
  physics with computers}},}\ }\href {\doibase 10.1007/BF02650179} {\bibfield
  {journal} {\bibinfo  {journal} {International Journal of Theoretical
  Physics}\ }\textbf {\bibinfo {volume} {21}},\ \bibinfo {pages} {467--488}
  (\bibinfo {year} {1982})}\BibitemShut {NoStop}%
\bibitem [{\citenamefont {Lloyd}(1996)}]{Lloyd1996}%
  \BibitemOpen
  \bibfield  {author} {\bibinfo {author} {\bibfnamefont {S.}~\bibnamefont
  {Lloyd}},\ }\bibfield  {title} {\enquote {\bibinfo {title} {{Universal
  quantum simulators}},}\ }\href {\doibase 10.1126/science.273.5278.1073}
  {\bibfield  {journal} {\bibinfo  {journal} {Science}\ }\textbf {\bibinfo
  {volume} {273}},\ \bibinfo {pages} {1073--1078} (\bibinfo {year}
  {1996})}\BibitemShut {NoStop}%
\bibitem [{Note1()}]{Note1}%
  \BibitemOpen
  \bibinfo {note} {Feynman's suggestion is also the basis of the broader
  research paradigm of quantum computing. The topic of quantum computing in
  general is extremely extensive, and an excellent introduction is given in the
  classic textbook \cite {Nielsen2010}}\BibitemShut {NoStop}%
\bibitem [{\citenamefont {Langford}\ \emph {et~al.}(2017)\citenamefont
  {Langford}, \citenamefont {Sagastizabal}, \citenamefont {Kounalakis},
  \citenamefont {Dickel}, \citenamefont {Bruno}, \citenamefont {Luthi},
  \citenamefont {Thoen}, \citenamefont {Endo},\ and\ \citenamefont
  {Dicarlo}}]{Langford2017}%
  \BibitemOpen
  \bibfield  {author} {\bibinfo {author} {\bibfnamefont {N.~K.}\ \bibnamefont
  {Langford}}, \bibinfo {author} {\bibfnamefont {R.}~\bibnamefont
  {Sagastizabal}}, \bibinfo {author} {\bibfnamefont {M.}~\bibnamefont
  {Kounalakis}}, \bibinfo {author} {\bibfnamefont {C.}~\bibnamefont {Dickel}},
  \bibinfo {author} {\bibfnamefont {A.}~\bibnamefont {Bruno}}, \bibinfo
  {author} {\bibfnamefont {F.}~\bibnamefont {Luthi}}, \bibinfo {author}
  {\bibfnamefont {D.~J.}\ \bibnamefont {Thoen}}, \bibinfo {author}
  {\bibfnamefont {A.}~\bibnamefont {Endo}}, \ and\ \bibinfo {author}
  {\bibfnamefont {L.}~\bibnamefont {Dicarlo}},\ }\bibfield  {title} {\enquote
  {\bibinfo {title} {{Experimentally simulating the dynamics of quantum light
  and matter at deep-strong coupling}},}\ }\href {\doibase
  10.1038/s41467-017-01061-x} {\bibfield  {journal} {\bibinfo  {journal}
  {Nature Communications}\ }\textbf {\bibinfo {volume} {8}},\ \bibinfo {pages}
  {1715} (\bibinfo {year} {2017})}\BibitemShut {NoStop}%
\bibitem [{\citenamefont {Lamata}\ \emph {et~al.}(2018)\citenamefont {Lamata},
  \citenamefont {Parra-Rodriguez}, \citenamefont {Sanz},\ and\ \citenamefont
  {Solano}}]{Lamata2018}%
  \BibitemOpen
  \bibfield  {author} {\bibinfo {author} {\bibfnamefont {L.}~\bibnamefont
  {Lamata}}, \bibinfo {author} {\bibfnamefont {A.}~\bibnamefont
  {Parra-Rodriguez}}, \bibinfo {author} {\bibfnamefont {M.}~\bibnamefont
  {Sanz}}, \ and\ \bibinfo {author} {\bibfnamefont {E.}~\bibnamefont
  {Solano}},\ }\bibfield  {title} {\enquote {\bibinfo {title} {{Digital-analog
  quantum simulations with superconducting circuits}},}\ }\href {\doibase
  10.1080/23746149.2018.1457981} {\bibfield  {journal} {\bibinfo  {journal}
  {Advances in Physics: X}\ }\textbf {\bibinfo {volume} {3}},\ \bibinfo {pages}
  {1457981} (\bibinfo {year} {2018})}\BibitemShut {NoStop}%
\bibitem [{\citenamefont {Wendin}(2017)}]{Wendin2017}%
  \BibitemOpen
  \bibfield  {author} {\bibinfo {author} {\bibfnamefont {G.}~\bibnamefont
  {Wendin}},\ }\bibfield  {title} {\enquote {\bibinfo {title} {{Quantum
  information processing with superconducting circuits: A review}},}\ }\href
  {\doibase 10.1088/1361-6633/aa7e1a} {\bibfield  {journal} {\bibinfo
  {journal} {Reports on Progress in Physics}\ }\textbf {\bibinfo {volume}
  {80}},\ \bibinfo {pages} {106001} (\bibinfo {year} {2017})}\BibitemShut
  {NoStop}%
\bibitem [{\citenamefont {Martinis}, \citenamefont {Devoret},\ and\
  \citenamefont {Clarke}(2020)}]{Martinis2020}%
  \BibitemOpen
  \bibfield  {author} {\bibinfo {author} {\bibfnamefont {J.~M.}\ \bibnamefont
  {Martinis}}, \bibinfo {author} {\bibfnamefont {M.~H.}\ \bibnamefont
  {Devoret}}, \ and\ \bibinfo {author} {\bibfnamefont {J.}~\bibnamefont
  {Clarke}},\ }\bibfield  {title} {\enquote {\bibinfo {title} {{Quantum
  Josephson junction circuits and the dawn of artificial atoms}},}\ }\href
  {\doibase 10.1038/s41567-020-0829-5} {\bibfield  {journal} {\bibinfo
  {journal} {Nature Physics}\ } (\bibinfo {year} {2020}),\
  10.1038/s41567-020-0829-5}\BibitemShut {NoStop}%
\bibitem [{\citenamefont {Carusotto}\ \emph {et~al.}(2020)\citenamefont
  {Carusotto}, \citenamefont {Houck}, \citenamefont {Koll}, \citenamefont
  {Roushan}, \citenamefont {Schuster}, \citenamefont {Simon}, \citenamefont
  {Koll{\'{a}}r}, \citenamefont {Roushan}, \citenamefont {Schuster},\ and\
  \citenamefont {Simon}}]{Carusotto2019}%
  \BibitemOpen
  \bibfield  {author} {\bibinfo {author} {\bibfnamefont {I.}~\bibnamefont
  {Carusotto}}, \bibinfo {author} {\bibfnamefont {A.~A.}\ \bibnamefont
  {Houck}}, \bibinfo {author} {\bibfnamefont {A.~J.}\ \bibnamefont {Koll}},
  \bibinfo {author} {\bibfnamefont {P.}~\bibnamefont {Roushan}}, \bibinfo
  {author} {\bibfnamefont {D.~I.}\ \bibnamefont {Schuster}}, \bibinfo {author}
  {\bibfnamefont {J.}~\bibnamefont {Simon}}, \bibinfo {author} {\bibfnamefont
  {A.~J.}\ \bibnamefont {Koll{\'{a}}r}}, \bibinfo {author} {\bibfnamefont
  {P.}~\bibnamefont {Roushan}}, \bibinfo {author} {\bibfnamefont {D.~I.}\
  \bibnamefont {Schuster}}, \ and\ \bibinfo {author} {\bibfnamefont
  {J.}~\bibnamefont {Simon}},\ }\bibfield  {title} {\enquote {\bibinfo {title}
  {{Photonic Materials in Circuit Quantum Electrodynamics}},}\ }\href {\doibase
  10.1038/s41567-020-0815-y} {\bibfield  {journal} {\bibinfo  {journal} {Nature
  Physics}\ }\textbf {\bibinfo {volume} {16}},\ \bibinfo {pages} {268}
  (\bibinfo {year} {2020})}\BibitemShut {NoStop}%
\bibitem [{\citenamefont {Blais}, \citenamefont {Girvin},\ and\ \citenamefont
  {Oliver}(2020)}]{Blais2020}%
  \BibitemOpen
  \bibfield  {author} {\bibinfo {author} {\bibfnamefont {A.}~\bibnamefont
  {Blais}}, \bibinfo {author} {\bibfnamefont {S.~M.}\ \bibnamefont {Girvin}}, \
  and\ \bibinfo {author} {\bibfnamefont {W.~D.}\ \bibnamefont {Oliver}},\
  }\bibfield  {title} {\enquote {\bibinfo {title} {{Quantum information
  processing and quantum optics with circuit quantum electrodynamics}},}\
  }\href {\doibase 10.1038/s41567-020-0806-z} {\bibfield  {journal} {\bibinfo
  {journal} {Nature Physics}\ } (\bibinfo {year} {2020}),\
  10.1038/s41567-020-0806-z}\BibitemShut {NoStop}%
\bibitem [{\citenamefont {Kjaergaard}\ \emph {et~al.}(2020)\citenamefont
  {Kjaergaard}, \citenamefont {Schwartz}, \citenamefont {Braum{\"{u}}ller},
  \citenamefont {Krantz}, \citenamefont {Wang}, \citenamefont {Gustavsson},\
  and\ \citenamefont {Oliver}}]{Kjaergaard2019}%
  \BibitemOpen
  \bibfield  {author} {\bibinfo {author} {\bibfnamefont {M.}~\bibnamefont
  {Kjaergaard}}, \bibinfo {author} {\bibfnamefont {M.~E.}\ \bibnamefont
  {Schwartz}}, \bibinfo {author} {\bibfnamefont {J.}~\bibnamefont
  {Braum{\"{u}}ller}}, \bibinfo {author} {\bibfnamefont {P.}~\bibnamefont
  {Krantz}}, \bibinfo {author} {\bibfnamefont {J.~I.-J.}\ \bibnamefont {Wang}},
  \bibinfo {author} {\bibfnamefont {S.}~\bibnamefont {Gustavsson}}, \ and\
  \bibinfo {author} {\bibfnamefont {W.~D.}\ \bibnamefont {Oliver}},\ }\bibfield
   {title} {\enquote {\bibinfo {title} {{Superconducting Qubits: Current State
  of Play}},}\ }\href {http://arxiv.org/abs/1905.13641} {\bibfield  {journal}
  {\bibinfo  {journal} {Annual Review of Condensed Matter Physics}\ }\textbf
  {\bibinfo {volume} {11}},\ \bibinfo {pages} {369} (\bibinfo {year}
  {2020})}\BibitemShut {NoStop}%
\bibitem [{\citenamefont {Krantz}\ \emph {et~al.}(2019)\citenamefont {Krantz},
  \citenamefont {Kjaergaard}, \citenamefont {Yan}, \citenamefont {Orlando},
  \citenamefont {Gustavsson},\ and\ \citenamefont {Oliver}}]{Krantz2019}%
  \BibitemOpen
  \bibfield  {author} {\bibinfo {author} {\bibfnamefont {P.}~\bibnamefont
  {Krantz}}, \bibinfo {author} {\bibfnamefont {M.}~\bibnamefont {Kjaergaard}},
  \bibinfo {author} {\bibfnamefont {F.}~\bibnamefont {Yan}}, \bibinfo {author}
  {\bibfnamefont {T.~P.}\ \bibnamefont {Orlando}}, \bibinfo {author}
  {\bibfnamefont {S.}~\bibnamefont {Gustavsson}}, \ and\ \bibinfo {author}
  {\bibfnamefont {W.~D.}\ \bibnamefont {Oliver}},\ }\bibfield  {title}
  {\enquote {\bibinfo {title} {{A quantum engineer's guide to superconducting
  qubits}},}\ }\href@noop {} {\bibfield  {journal} {\bibinfo  {journal}
  {Applied Physics Reviews}\ }\textbf {\bibinfo {volume} {6}},\ \bibinfo
  {pages} {021318} (\bibinfo {year} {2019})}\BibitemShut {NoStop}%
\bibitem [{\citenamefont {Koch}\ \emph {et~al.}(2007)\citenamefont {Koch},
  \citenamefont {Yu}, \citenamefont {Gambetta}, \citenamefont {Houck},
  \citenamefont {Schuster}, \citenamefont {Majer}, \citenamefont {Blais},
  \citenamefont {Devoret}, \citenamefont {Girvin},\ and\ \citenamefont
  {Schoelkopf}}]{Koch2007}%
  \BibitemOpen
  \bibfield  {author} {\bibinfo {author} {\bibfnamefont {J.}~\bibnamefont
  {Koch}}, \bibinfo {author} {\bibfnamefont {T.~M.}\ \bibnamefont {Yu}},
  \bibinfo {author} {\bibfnamefont {J.}~\bibnamefont {Gambetta}}, \bibinfo
  {author} {\bibfnamefont {A.~A.}\ \bibnamefont {Houck}}, \bibinfo {author}
  {\bibfnamefont {D.~I.}\ \bibnamefont {Schuster}}, \bibinfo {author}
  {\bibfnamefont {J.}~\bibnamefont {Majer}}, \bibinfo {author} {\bibfnamefont
  {A.}~\bibnamefont {Blais}}, \bibinfo {author} {\bibfnamefont {M.~H.}\
  \bibnamefont {Devoret}}, \bibinfo {author} {\bibfnamefont {S.~M.}\
  \bibnamefont {Girvin}}, \ and\ \bibinfo {author} {\bibfnamefont {R.~J.}\
  \bibnamefont {Schoelkopf}},\ }\bibfield  {title} {\enquote {\bibinfo {title}
  {{Charge-insensitive qubit design derived from the Cooper pair box}},}\
  }\href {\doibase 10.1103/PhysRevA.76.042319} {\bibfield  {journal} {\bibinfo
  {journal} {Physical Review A}\ }\textbf {\bibinfo {volume} {76}},\ \bibinfo
  {pages} {42319} (\bibinfo {year} {2007})}\BibitemShut {NoStop}%
\bibitem [{\citenamefont {Houck}\ \emph {et~al.}(2009)\citenamefont {Houck},
  \citenamefont {Koch}, \citenamefont {Devoret}, \citenamefont {Girvin},\ and\
  \citenamefont {Schoelkopf}}]{Houck2009}%
  \BibitemOpen
  \bibfield  {author} {\bibinfo {author} {\bibfnamefont {A.~A.}\ \bibnamefont
  {Houck}}, \bibinfo {author} {\bibfnamefont {J.}~\bibnamefont {Koch}},
  \bibinfo {author} {\bibfnamefont {M.~H.}\ \bibnamefont {Devoret}}, \bibinfo
  {author} {\bibfnamefont {S.~M.}\ \bibnamefont {Girvin}}, \ and\ \bibinfo
  {author} {\bibfnamefont {R.~J.}\ \bibnamefont {Schoelkopf}},\ }\bibfield
  {title} {\enquote {\bibinfo {title} {{Life after charge noise: Recent results
  with transmon qubits}},}\ }\href {\doibase 10.1007/s11128-009-0100-6}
  {\bibfield  {journal} {\bibinfo  {journal} {Quantum Information Processing}\
  }\textbf {\bibinfo {volume} {8}},\ \bibinfo {pages} {105--115} (\bibinfo
  {year} {2009})}\BibitemShut {NoStop}%
\bibitem [{Note2()}]{Note2}%
  \BibitemOpen
  \bibinfo {note} {An alternative qubit design, the capacitively-shunted flux
  qubit \cite {Mooij1999,Yan2016}, has much larger non-linearity with similar
  single-qubit coherence times to transmons. However, presently these flux
  qubits cannot be coupled with the scalability and flexibility of transmons
  without introducing additional noise and thus reducing
  coherence.}\BibitemShut {Stop}%
\bibitem [{\citenamefont {Neumeier}, \citenamefont {Leib},\ and\ \citenamefont
  {Hartmann}(2013)}]{Neumeier2013}%
  \BibitemOpen
  \bibfield  {author} {\bibinfo {author} {\bibfnamefont {L.}~\bibnamefont
  {Neumeier}}, \bibinfo {author} {\bibfnamefont {M.}~\bibnamefont {Leib}}, \
  and\ \bibinfo {author} {\bibfnamefont {M.~J.}\ \bibnamefont {Hartmann}},\
  }\bibfield  {title} {\enquote {\bibinfo {title} {{Single-photon transistor in
  circuit quantum electrodynamics}},}\ }\href {\doibase
  10.1103/PhysRevLett.111.063601} {\bibfield  {journal} {\bibinfo  {journal}
  {Physical Review Letters}\ }\textbf {\bibinfo {volume} {111}},\ \bibinfo
  {pages} {063601} (\bibinfo {year} {2013})}\BibitemShut {NoStop}%
\bibitem [{\citenamefont {Jin}\ \emph {et~al.}(2013)\citenamefont {Jin},
  \citenamefont {Rossini}, \citenamefont {Fazio}, \citenamefont {Leib},\ and\
  \citenamefont {Hartmann}}]{Jin2013}%
  \BibitemOpen
  \bibfield  {author} {\bibinfo {author} {\bibfnamefont {J.}~\bibnamefont
  {Jin}}, \bibinfo {author} {\bibfnamefont {D.}~\bibnamefont {Rossini}},
  \bibinfo {author} {\bibfnamefont {R.}~\bibnamefont {Fazio}}, \bibinfo
  {author} {\bibfnamefont {M.}~\bibnamefont {Leib}}, \ and\ \bibinfo {author}
  {\bibfnamefont {M.~J.}\ \bibnamefont {Hartmann}},\ }\bibfield  {title}
  {\enquote {\bibinfo {title} {{Photon solid phases in driven arrays of
  nonlinearly coupled cavities}},}\ }\href {\doibase
  10.1103/PhysRevLett.110.163605} {\bibfield  {journal} {\bibinfo  {journal}
  {Physical Review Letters}\ }\textbf {\bibinfo {volume} {110}},\ \bibinfo
  {pages} {163605} (\bibinfo {year} {2013})}\BibitemShut {NoStop}%
\bibitem [{\citenamefont {Collodo}\ \emph {et~al.}(2019)\citenamefont
  {Collodo}, \citenamefont {Poto{\v{c}}nik}, \citenamefont {Gasparinetti},
  \citenamefont {Besse}, \citenamefont {Pechal}, \citenamefont {Sameti},
  \citenamefont {Hartmann}, \citenamefont {Wallraff},\ and\ \citenamefont
  {Eichler}}]{Collodo2019}%
  \BibitemOpen
  \bibfield  {author} {\bibinfo {author} {\bibfnamefont {M.~C.}\ \bibnamefont
  {Collodo}}, \bibinfo {author} {\bibfnamefont {A.}~\bibnamefont
  {Poto{\v{c}}nik}}, \bibinfo {author} {\bibfnamefont {S.}~\bibnamefont
  {Gasparinetti}}, \bibinfo {author} {\bibfnamefont {J.~C.}\ \bibnamefont
  {Besse}}, \bibinfo {author} {\bibfnamefont {M.}~\bibnamefont {Pechal}},
  \bibinfo {author} {\bibfnamefont {M.}~\bibnamefont {Sameti}}, \bibinfo
  {author} {\bibfnamefont {M.~J.}\ \bibnamefont {Hartmann}}, \bibinfo {author}
  {\bibfnamefont {A.}~\bibnamefont {Wallraff}}, \ and\ \bibinfo {author}
  {\bibfnamefont {C.}~\bibnamefont {Eichler}},\ }\bibfield  {title} {\enquote
  {\bibinfo {title} {{Observation of the Crossover from Photon Ordering to
  Delocalization in Tunably Coupled Resonators}},}\ }\href {\doibase
  10.1103/PhysRevLett.122.183601} {\bibfield  {journal} {\bibinfo  {journal}
  {Physical Review Letters}\ }\textbf {\bibinfo {volume} {122}},\ \bibinfo
  {pages} {183601} (\bibinfo {year} {2019})}\BibitemShut {NoStop}%
\bibitem [{Note3()}]{Note3}%
  \BibitemOpen
  \bibinfo {note} {With one caveat -- one must be careful not to create closed
  superconducting loops, which lead to flux quantization and make the system
  more sensitive to magnetic flux noise. This can be avoided by introducing a
  capacitor into the edges, such that the loop is broken.}\BibitemShut {Stop}%
\bibitem [{\citenamefont {Fisher}\ \emph {et~al.}(1989)\citenamefont {Fisher},
  \citenamefont {Weichman}, \citenamefont {Grinstein},\ and\ \citenamefont
  {Fisher}}]{Fisher1989}%
  \BibitemOpen
  \bibfield  {author} {\bibinfo {author} {\bibfnamefont {M.~P.~A.}\
  \bibnamefont {Fisher}}, \bibinfo {author} {\bibfnamefont {P.~B.}\
  \bibnamefont {Weichman}}, \bibinfo {author} {\bibfnamefont {G.}~\bibnamefont
  {Grinstein}}, \ and\ \bibinfo {author} {\bibfnamefont {D.~S.}\ \bibnamefont
  {Fisher}},\ }\bibfield  {title} {\enquote {\bibinfo {title} {{Boson
  localization and the superfluid-insulator transition}},}\ }\href {\doibase
  10.1103/PhysRevB.40.546} {\bibfield  {journal} {\bibinfo  {journal} {Physical
  Review B}\ }\textbf {\bibinfo {volume} {40}},\ \bibinfo {pages} {546}
  (\bibinfo {year} {1989})}\BibitemShut {NoStop}%
\bibitem [{\citenamefont {Sachdev}(2011)}]{Sachdev2011}%
  \BibitemOpen
  \bibfield  {author} {\bibinfo {author} {\bibfnamefont {S.}~\bibnamefont
  {Sachdev}},\ }\href@noop {} {\emph {\bibinfo {title} {{Quantum Phase
  Transitions}}}}\ (\bibinfo  {publisher} {Cambridge University Press},\
  \bibinfo {year} {2011})\BibitemShut {NoStop}%
\bibitem [{\citenamefont {Rossini}\ and\ \citenamefont
  {Fazio}(2012)}]{Rossini2012}%
  \BibitemOpen
  \bibfield  {author} {\bibinfo {author} {\bibfnamefont {D.}~\bibnamefont
  {Rossini}}\ and\ \bibinfo {author} {\bibfnamefont {R.}~\bibnamefont
  {Fazio}},\ }\bibfield  {title} {\enquote {\bibinfo {title} {{Phase diagram of
  the extended Bose Hubbard model}},}\ }\href {\doibase
  10.1088/1367-2630/14/6/065012} {\bibfield  {journal} {\bibinfo  {journal}
  {New Journal of Physics}\ }\textbf {\bibinfo {volume} {14}},\ \bibinfo
  {pages} {65012} (\bibinfo {year} {2012})}\BibitemShut {NoStop}%
\bibitem [{\citenamefont {Kitaev}(2002)}]{Kitaev2002}%
  \BibitemOpen
  \bibfield  {author} {\bibinfo {author} {\bibfnamefont {A.}~\bibnamefont
  {Kitaev}},\ }\bibfield  {title} {\enquote {\bibinfo {title} {{Fault-tolerant
  quantum computation by anyons}},}\ }\href@noop {} {\bibfield  {journal}
  {\bibinfo  {journal} {Annals of Physics}\ }\textbf {\bibinfo {volume}
  {303}},\ \bibinfo {pages} {2--30} (\bibinfo {year} {2002})}\BibitemShut
  {NoStop}%
\bibitem [{\citenamefont {Kitaev}\ and\ \citenamefont
  {Laumann}(2009)}]{Kitaev2009}%
  \BibitemOpen
  \bibfield  {author} {\bibinfo {author} {\bibfnamefont {A.}~\bibnamefont
  {Kitaev}}\ and\ \bibinfo {author} {\bibfnamefont {C.}~\bibnamefont
  {Laumann}},\ }\bibfield  {title} {\enquote {\bibinfo {title} {{Topological
  phases and quantum computation}},}\ }\href {http://arxiv.org/abs/0904.2771}
  {\  (\bibinfo {year} {2009})},\ \Eprint {http://arxiv.org/abs/0904.2771}
  {arXiv:0904.2771} \BibitemShut {NoStop}%
\bibitem [{\citenamefont {Fowler}\ \emph {et~al.}(2012)\citenamefont {Fowler},
  \citenamefont {Mariantoni}, \citenamefont {Martinis},\ and\ \citenamefont
  {Cleland}}]{Fowler2012}%
  \BibitemOpen
  \bibfield  {author} {\bibinfo {author} {\bibfnamefont {A.~G.}\ \bibnamefont
  {Fowler}}, \bibinfo {author} {\bibfnamefont {M.}~\bibnamefont {Mariantoni}},
  \bibinfo {author} {\bibfnamefont {J.~M.}\ \bibnamefont {Martinis}}, \ and\
  \bibinfo {author} {\bibfnamefont {A.~N.}\ \bibnamefont {Cleland}},\
  }\bibfield  {title} {\enquote {\bibinfo {title} {{Surface codes: Towards
  practical large-scale quantum computation}},}\ }\href {\doibase
  10.1103/PhysRevA.86.032324} {\bibfield  {journal} {\bibinfo  {journal}
  {Physical Review A}\ }\textbf {\bibinfo {volume} {86}},\ \bibinfo {pages}
  {032324} (\bibinfo {year} {2012})}\BibitemShut {NoStop}%
\bibitem [{\citenamefont {Sameti}\ \emph {et~al.}(2017)\citenamefont {Sameti},
  \citenamefont {Poto{\v{c}}nik}, \citenamefont {Browne}, \citenamefont
  {Wallraff},\ and\ \citenamefont {Hartmann}}]{Sameti2017}%
  \BibitemOpen
  \bibfield  {author} {\bibinfo {author} {\bibfnamefont {M.}~\bibnamefont
  {Sameti}}, \bibinfo {author} {\bibfnamefont {A.}~\bibnamefont
  {Poto{\v{c}}nik}}, \bibinfo {author} {\bibfnamefont {D.~E.}\ \bibnamefont
  {Browne}}, \bibinfo {author} {\bibfnamefont {A.}~\bibnamefont {Wallraff}}, \
  and\ \bibinfo {author} {\bibfnamefont {M.~J.}\ \bibnamefont {Hartmann}},\
  }\bibfield  {title} {\enquote {\bibinfo {title} {{Superconducting quantum
  simulator for topological order and the toric code}},}\ }\href {\doibase
  10.1103/PhysRevA.95.042330} {\bibfield  {journal} {\bibinfo  {journal}
  {Physical Review A}\ }\textbf {\bibinfo {volume} {95}},\ \bibinfo {pages}
  {042330} (\bibinfo {year} {2017})}\BibitemShut {NoStop}%
\bibitem [{\citenamefont {Niskanen}\ \emph {et~al.}(2007)\citenamefont
  {Niskanen}, \citenamefont {Harrabi}, \citenamefont {Yoshihara}, \citenamefont
  {Nakamura}, \citenamefont {Lloyd},\ and\ \citenamefont
  {Tsai}}]{Niskanen2007}%
  \BibitemOpen
  \bibfield  {author} {\bibinfo {author} {\bibfnamefont {A.~O.}\ \bibnamefont
  {Niskanen}}, \bibinfo {author} {\bibfnamefont {K.}~\bibnamefont {Harrabi}},
  \bibinfo {author} {\bibfnamefont {F.}~\bibnamefont {Yoshihara}}, \bibinfo
  {author} {\bibfnamefont {Y.}~\bibnamefont {Nakamura}}, \bibinfo {author}
  {\bibfnamefont {S.}~\bibnamefont {Lloyd}}, \ and\ \bibinfo {author}
  {\bibfnamefont {J.~S.}\ \bibnamefont {Tsai}},\ }\bibfield  {title} {\enquote
  {\bibinfo {title} {{Quantum coherent tunable coupling of superconducting
  qubits}},}\ }\href {\doibase 10.1126/science.1141324} {\bibfield  {journal}
  {\bibinfo  {journal} {Science}\ }\textbf {\bibinfo {volume} {316}},\ \bibinfo
  {pages} {723--726} (\bibinfo {year} {2007})}\BibitemShut {NoStop}%
\bibitem [{\citenamefont {Roushan}\ \emph {et~al.}(2017)\citenamefont
  {Roushan}, \citenamefont {Neill}, \citenamefont {Megrant}, \citenamefont
  {Chen}, \citenamefont {Babbush}, \citenamefont {Barends}, \citenamefont
  {Campbell}, \citenamefont {Chen}, \citenamefont {Chiaro}, \citenamefont
  {Dunsworth}, \citenamefont {Fowler}, \citenamefont {Jeffrey}, \citenamefont
  {Kelly}, \citenamefont {Lucero}, \citenamefont {Mutus}, \citenamefont
  {O'Malley}, \citenamefont {Neeley}, \citenamefont {Quintana}, \citenamefont
  {Sank}, \citenamefont {Vainsencher}, \citenamefont {Wenner}, \citenamefont
  {White}, \citenamefont {Kapit}, \citenamefont {Neven},\ and\ \citenamefont
  {Martinis}}]{Roushan2017}%
  \BibitemOpen
  \bibfield  {author} {\bibinfo {author} {\bibfnamefont {P.}~\bibnamefont
  {Roushan}}, \bibinfo {author} {\bibfnamefont {C.}~\bibnamefont {Neill}},
  \bibinfo {author} {\bibfnamefont {A.}~\bibnamefont {Megrant}}, \bibinfo
  {author} {\bibfnamefont {Y.}~\bibnamefont {Chen}}, \bibinfo {author}
  {\bibfnamefont {R.}~\bibnamefont {Babbush}}, \bibinfo {author} {\bibfnamefont
  {R.}~\bibnamefont {Barends}}, \bibinfo {author} {\bibfnamefont
  {B.}~\bibnamefont {Campbell}}, \bibinfo {author} {\bibfnamefont
  {Z.}~\bibnamefont {Chen}}, \bibinfo {author} {\bibfnamefont {B.}~\bibnamefont
  {Chiaro}}, \bibinfo {author} {\bibfnamefont {A.}~\bibnamefont {Dunsworth}},
  \bibinfo {author} {\bibfnamefont {A.}~\bibnamefont {Fowler}}, \bibinfo
  {author} {\bibfnamefont {E.}~\bibnamefont {Jeffrey}}, \bibinfo {author}
  {\bibfnamefont {J.}~\bibnamefont {Kelly}}, \bibinfo {author} {\bibfnamefont
  {E.}~\bibnamefont {Lucero}}, \bibinfo {author} {\bibfnamefont
  {J.}~\bibnamefont {Mutus}}, \bibinfo {author} {\bibfnamefont {P.~J.}\
  \bibnamefont {O'Malley}}, \bibinfo {author} {\bibfnamefont {M.}~\bibnamefont
  {Neeley}}, \bibinfo {author} {\bibfnamefont {C.}~\bibnamefont {Quintana}},
  \bibinfo {author} {\bibfnamefont {D.}~\bibnamefont {Sank}}, \bibinfo {author}
  {\bibfnamefont {A.}~\bibnamefont {Vainsencher}}, \bibinfo {author}
  {\bibfnamefont {J.}~\bibnamefont {Wenner}}, \bibinfo {author} {\bibfnamefont
  {T.}~\bibnamefont {White}}, \bibinfo {author} {\bibfnamefont
  {E.}~\bibnamefont {Kapit}}, \bibinfo {author} {\bibfnamefont
  {H.}~\bibnamefont {Neven}}, \ and\ \bibinfo {author} {\bibfnamefont
  {J.}~\bibnamefont {Martinis}},\ }\bibfield  {title} {\enquote {\bibinfo
  {title} {{Chiral ground-state currents of interacting photons in a synthetic
  magnetic field}},}\ }\href {\doibase 10.1038/nphys3930} {\bibfield  {journal}
  {\bibinfo  {journal} {Nature Physics}\ }\textbf {\bibinfo {volume} {13}},\
  \bibinfo {pages} {146--151} (\bibinfo {year} {2017})}\BibitemShut {NoStop}%
\bibitem [{\citenamefont {Braum{\"u}ller}\ \emph {et~al.}(2017)\citenamefont
  {Braum{\"u}ller}, \citenamefont {Marthaler}, \citenamefont {Schneider},
  \citenamefont {Stehli}, \citenamefont {Rotzinger}, \citenamefont {Weides},\
  and\ \citenamefont {Ustinov}}]{Braumuller:2017zl}%
  \BibitemOpen
  \bibfield  {author} {\bibinfo {author} {\bibfnamefont {J.}~\bibnamefont
  {Braum{\"u}ller}}, \bibinfo {author} {\bibfnamefont {M.}~\bibnamefont
  {Marthaler}}, \bibinfo {author} {\bibfnamefont {A.}~\bibnamefont
  {Schneider}}, \bibinfo {author} {\bibfnamefont {A.}~\bibnamefont {Stehli}},
  \bibinfo {author} {\bibfnamefont {H.}~\bibnamefont {Rotzinger}}, \bibinfo
  {author} {\bibfnamefont {M.}~\bibnamefont {Weides}}, \ and\ \bibinfo {author}
  {\bibfnamefont {A.~V.}\ \bibnamefont {Ustinov}},\ }\bibfield  {title}
  {\enquote {\bibinfo {title} {Analog quantum simulation of the rabi model in
  the ultra-strong coupling regime},}\ }\href {\doibase
  10.1038/s41467-017-00894-w} {\bibfield  {journal} {\bibinfo  {journal}
  {Nature Communications}\ }\textbf {\bibinfo {volume} {8}},\ \bibinfo {pages}
  {779} (\bibinfo {year} {2017})}\BibitemShut {NoStop}%
\bibitem [{\citenamefont {Oka}\ and\ \citenamefont {Kitamura}(2019)}]{Oka2019}%
  \BibitemOpen
  \bibfield  {author} {\bibinfo {author} {\bibfnamefont {T.}~\bibnamefont
  {Oka}}\ and\ \bibinfo {author} {\bibfnamefont {S.}~\bibnamefont {Kitamura}},\
  }\bibfield  {title} {\enquote {\bibinfo {title} {{Floquet Engineering of
  Quantum Materials}},}\ }\href {\doibase
  10.1146/annurev-conmatphys-031218-013423} {\bibfield  {journal} {\bibinfo
  {journal} {Annual Review of Condensed Matter Physics}\ }\textbf {\bibinfo
  {volume} {10}},\ \bibinfo {pages} {387--408} (\bibinfo {year}
  {2019})}\BibitemShut {NoStop}%
\bibitem [{\citenamefont {Sameti}\ and\ \citenamefont
  {Hartmann}(2019)}]{Sameti2019}%
  \BibitemOpen
  \bibfield  {author} {\bibinfo {author} {\bibfnamefont {M.}~\bibnamefont
  {Sameti}}\ and\ \bibinfo {author} {\bibfnamefont {M.~J.}\ \bibnamefont
  {Hartmann}},\ }\bibfield  {title} {\enquote {\bibinfo {title} {{Floquet
  engineering in superconducting circuits: From arbitrary spin-spin
  interactions to the Kitaev honeycomb model}},}\ }\href {\doibase
  10.1103/PhysRevA.99.012333} {\bibfield  {journal} {\bibinfo  {journal}
  {Physical Review A}\ }\textbf {\bibinfo {volume} {99}},\ \bibinfo {pages}
  {012333} (\bibinfo {year} {2019})}\BibitemShut {NoStop}%
\bibitem [{\citenamefont {Peierls}(1933)}]{Peierls1933}%
  \BibitemOpen
  \bibfield  {author} {\bibinfo {author} {\bibfnamefont {R.}~\bibnamefont
  {Peierls}},\ }\bibfield  {title} {\enquote {\bibinfo {title} {{Zur Theorie
  des Diamagnetismus von Leitungselektronen.}}}\ }\href@noop {} {\bibfield
  {journal} {\bibinfo  {journal} {Zeitschrift f{\"{u}}r Physik}\ }\textbf
  {\bibinfo {volume} {80}},\ \bibinfo {pages} {763--791} (\bibinfo {year}
  {1933})}\BibitemShut {NoStop}%
\bibitem [{\citenamefont {Fang}, \citenamefont {Yu},\ and\ \citenamefont
  {Fan}(2012)}]{Fang2012}%
  \BibitemOpen
  \bibfield  {author} {\bibinfo {author} {\bibfnamefont {K.}~\bibnamefont
  {Fang}}, \bibinfo {author} {\bibfnamefont {Z.}~\bibnamefont {Yu}}, \ and\
  \bibinfo {author} {\bibfnamefont {S.}~\bibnamefont {Fan}},\ }\bibfield
  {title} {\enquote {\bibinfo {title} {{Realizing effective magnetic field for
  photons by controlling the phase of dynamic modulation}},}\ }\href {\doibase
  10.1038/nphoton.2012.236} {\bibfield  {journal} {\bibinfo  {journal} {Nature
  Photonics}\ }\textbf {\bibinfo {volume} {6}},\ \bibinfo {pages} {782--787}
  (\bibinfo {year} {2012})}\BibitemShut {NoStop}%
\bibitem [{\citenamefont {Koch}\ \emph {et~al.}(2010)\citenamefont {Koch},
  \citenamefont {Houck}, \citenamefont {Hur},\ and\ \citenamefont
  {Girvin}}]{Koch2010}%
  \BibitemOpen
  \bibfield  {author} {\bibinfo {author} {\bibfnamefont {J.}~\bibnamefont
  {Koch}}, \bibinfo {author} {\bibfnamefont {A.~A.}\ \bibnamefont {Houck}},
  \bibinfo {author} {\bibfnamefont {K.~L.}\ \bibnamefont {Hur}}, \ and\
  \bibinfo {author} {\bibfnamefont {S.~M.}\ \bibnamefont {Girvin}},\ }\bibfield
   {title} {\enquote {\bibinfo {title} {{Time-reversal-symmetry breaking in
  circuit-QED-based photon lattices}},}\ }\href {\doibase
  10.1103/PhysRevA.82.043811} {\bibfield  {journal} {\bibinfo  {journal}
  {Physical Review A}\ }\textbf {\bibinfo {volume} {82}},\ \bibinfo {pages}
  {043811} (\bibinfo {year} {2010})}\BibitemShut {NoStop}%
\bibitem [{\citenamefont {M{\"{u}}ller}\ \emph {et~al.}(2018)\citenamefont
  {M{\"{u}}ller}, \citenamefont {Guan}, \citenamefont {Vogt}, \citenamefont
  {Cole},\ and\ \citenamefont {Stace}}]{Muller2018}%
  \BibitemOpen
  \bibfield  {author} {\bibinfo {author} {\bibfnamefont {C.}~\bibnamefont
  {M{\"{u}}ller}}, \bibinfo {author} {\bibfnamefont {S.}~\bibnamefont {Guan}},
  \bibinfo {author} {\bibfnamefont {N.}~\bibnamefont {Vogt}}, \bibinfo {author}
  {\bibfnamefont {J.~H.}\ \bibnamefont {Cole}}, \ and\ \bibinfo {author}
  {\bibfnamefont {T.~M.}\ \bibnamefont {Stace}},\ }\bibfield  {title} {\enquote
  {\bibinfo {title} {{Passive On-Chip Superconducting Circulator Using a Ring
  of Tunnel Junctions}},}\ }\href {\doibase 10.1103/PhysRevLett.120.213602}
  {\bibfield  {journal} {\bibinfo  {journal} {Physical Review Letters}\
  }\textbf {\bibinfo {volume} {120}},\ \bibinfo {pages} {213602} (\bibinfo
  {year} {2018})}\BibitemShut {NoStop}%
\bibitem [{\citenamefont {Ozawa}\ \emph {et~al.}(2019)\citenamefont {Ozawa},
  \citenamefont {Price}, \citenamefont {Amo}, \citenamefont {Goldman},
  \citenamefont {Hafezi}, \citenamefont {Lu}, \citenamefont {Rechtsman},
  \citenamefont {Schuster}, \citenamefont {Simon}, \citenamefont {Zilberberg},\
  and\ \citenamefont {Carusotto}}]{Ozawa2019}%
  \BibitemOpen
  \bibfield  {author} {\bibinfo {author} {\bibfnamefont {T.}~\bibnamefont
  {Ozawa}}, \bibinfo {author} {\bibfnamefont {H.~M.}\ \bibnamefont {Price}},
  \bibinfo {author} {\bibfnamefont {A.}~\bibnamefont {Amo}}, \bibinfo {author}
  {\bibfnamefont {N.}~\bibnamefont {Goldman}}, \bibinfo {author} {\bibfnamefont
  {M.}~\bibnamefont {Hafezi}}, \bibinfo {author} {\bibfnamefont
  {L.}~\bibnamefont {Lu}}, \bibinfo {author} {\bibfnamefont {M.~C.}\
  \bibnamefont {Rechtsman}}, \bibinfo {author} {\bibfnamefont {D.}~\bibnamefont
  {Schuster}}, \bibinfo {author} {\bibfnamefont {J.}~\bibnamefont {Simon}},
  \bibinfo {author} {\bibfnamefont {O.}~\bibnamefont {Zilberberg}}, \ and\
  \bibinfo {author} {\bibfnamefont {I.}~\bibnamefont {Carusotto}},\ }\bibfield
  {title} {\enquote {\bibinfo {title} {{Topological photonics}},}\ }\href
  {\doibase 10.1038/nphoton.2014.248} {\bibfield  {journal} {\bibinfo
  {journal} {Reviews of Modern Physics}\ }\textbf {\bibinfo {volume} {91}},\
  \bibinfo {pages} {015006} (\bibinfo {year} {2019})}\BibitemShut {NoStop}%
\bibitem [{\citenamefont {Bloch}, \citenamefont {Dalibard},\ and\ \citenamefont
  {Zwerger}(2008)}]{Bloch2008}%
  \BibitemOpen
  \bibfield  {author} {\bibinfo {author} {\bibfnamefont {I.}~\bibnamefont
  {Bloch}}, \bibinfo {author} {\bibfnamefont {J.}~\bibnamefont {Dalibard}}, \
  and\ \bibinfo {author} {\bibfnamefont {W.}~\bibnamefont {Zwerger}},\
  }\bibfield  {title} {\enquote {\bibinfo {title} {{Many-body physics with
  ultracold gases}},}\ }\href {\doibase 10.1103/RevModPhys.80.885} {\bibfield
  {journal} {\bibinfo  {journal} {Reviews of Modern Physics}\ }\textbf
  {\bibinfo {volume} {80}},\ \bibinfo {pages} {885} (\bibinfo {year}
  {2008})}\BibitemShut {NoStop}%
\bibitem [{\citenamefont {Eisert}, \citenamefont {Friesdorf},\ and\
  \citenamefont {Gogolin}(2015)}]{Eisert2015}%
  \BibitemOpen
  \bibfield  {author} {\bibinfo {author} {\bibfnamefont {J.}~\bibnamefont
  {Eisert}}, \bibinfo {author} {\bibfnamefont {M.}~\bibnamefont {Friesdorf}}, \
  and\ \bibinfo {author} {\bibfnamefont {C.}~\bibnamefont {Gogolin}},\
  }\bibfield  {title} {\enquote {\bibinfo {title} {{Quantum many-body systems
  out of equilibrium}},}\ }\href {\doibase 10.1038/nphys3215} {\bibfield
  {journal} {\bibinfo  {journal} {Nature Physics}\ }\textbf {\bibinfo {volume}
  {11}},\ \bibinfo {pages} {124--130} (\bibinfo {year} {2015})}\BibitemShut
  {NoStop}%
\bibitem [{\citenamefont {Raftery}\ \emph {et~al.}(2014)\citenamefont
  {Raftery}, \citenamefont {Sadri}, \citenamefont {Schmidt}, \citenamefont
  {Tu¨reci},\ and\ \citenamefont {Houck}}]{Raftery2014}%
  \BibitemOpen
  \bibfield  {author} {\bibinfo {author} {\bibfnamefont {J.}~\bibnamefont
  {Raftery}}, \bibinfo {author} {\bibfnamefont {D.}~\bibnamefont {Sadri}},
  \bibinfo {author} {\bibfnamefont {S.}~\bibnamefont {Schmidt}}, \bibinfo
  {author} {\bibfnamefont {H.~E.}\ \bibnamefont {T\"{u}reci}}, \ and\ \bibinfo
  {author} {\bibfnamefont {A.~A.}\ \bibnamefont {Houck}},\ }\bibfield  {title}
  {\enquote {\bibinfo {title} {{Observation of a dissipation-induced classical
  to quantum transition}},}\ }\href {\doibase 10.1103/PhysRevX.4.031043}
  {\bibfield  {journal} {\bibinfo  {journal} {Physical Review X}\ }\textbf
  {\bibinfo {volume} {4}},\ \bibinfo {pages} {031043} (\bibinfo {year}
  {2014})}\BibitemShut {NoStop}%
\bibitem [{\citenamefont {Fitzpatrick}\ \emph {et~al.}(2017)\citenamefont
  {Fitzpatrick}, \citenamefont {Sundaresan}, \citenamefont {Li}, \citenamefont
  {Koch},\ and\ \citenamefont {Houck}}]{Fitzpatrick2017}%
  \BibitemOpen
  \bibfield  {author} {\bibinfo {author} {\bibfnamefont {M.}~\bibnamefont
  {Fitzpatrick}}, \bibinfo {author} {\bibfnamefont {N.~M.}\ \bibnamefont
  {Sundaresan}}, \bibinfo {author} {\bibfnamefont {A.~C.}\ \bibnamefont {Li}},
  \bibinfo {author} {\bibfnamefont {J.}~\bibnamefont {Koch}}, \ and\ \bibinfo
  {author} {\bibfnamefont {A.~A.}\ \bibnamefont {Houck}},\ }\bibfield  {title}
  {\enquote {\bibinfo {title} {{Observation of a dissipative phase transition
  in a one-dimensional circuit QED lattice}},}\ }\href {\doibase
  10.1103/PhysRevX.7.011016} {\bibfield  {journal} {\bibinfo  {journal}
  {Physical Review X}\ }\textbf {\bibinfo {volume} {7}},\ \bibinfo {pages}
  {011016} (\bibinfo {year} {2017})}\BibitemShut {NoStop}%
\bibitem [{\citenamefont {Hartmann}(2010)}]{Hartmann2010}%
  \BibitemOpen
  \bibfield  {author} {\bibinfo {author} {\bibfnamefont {M.~J.}\ \bibnamefont
  {Hartmann}},\ }\bibfield  {title} {\enquote {\bibinfo {title} {{Polariton
  crystallization in driven arrays of lossy nonlinear resonators}},}\ }\href
  {\doibase 10.1103/PhysRevLett.104.113601} {\bibfield  {journal} {\bibinfo
  {journal} {Physical Review Letters}\ }\textbf {\bibinfo {volume} {104}},\
  \bibinfo {pages} {113601} (\bibinfo {year} {2010})}\BibitemShut {NoStop}%
\bibitem [{\citenamefont {Kounalakis}\ \emph {et~al.}(2018)\citenamefont
  {Kounalakis}, \citenamefont {Dickel}, \citenamefont {Bruno}, \citenamefont
  {Langford},\ and\ \citenamefont {Steele}}]{Kounalakis2018}%
  \BibitemOpen
  \bibfield  {author} {\bibinfo {author} {\bibfnamefont {M.}~\bibnamefont
  {Kounalakis}}, \bibinfo {author} {\bibfnamefont {C.}~\bibnamefont {Dickel}},
  \bibinfo {author} {\bibfnamefont {A.}~\bibnamefont {Bruno}}, \bibinfo
  {author} {\bibfnamefont {N.~K.}\ \bibnamefont {Langford}}, \ and\ \bibinfo
  {author} {\bibfnamefont {G.~A.}\ \bibnamefont {Steele}},\ }\bibfield  {title}
  {\enquote {\bibinfo {title} {{Tuneable hopping and nonlinear cross-Kerr
  interactions in a high-coherence superconducting circuit}},}\ }\href
  {\doibase 10.1038/s41534-018-0088-9} {\bibfield  {journal} {\bibinfo
  {journal} {npj Quantum Information}\ }\textbf {\bibinfo {volume} {4}},\
  \bibinfo {pages} {38} (\bibinfo {year} {2018})}\BibitemShut {NoStop}%
\bibitem [{\citenamefont {Leib}\ and\ \citenamefont
  {Hartmann}(2014)}]{Leib2014}%
  \BibitemOpen
  \bibfield  {author} {\bibinfo {author} {\bibfnamefont {M.}~\bibnamefont
  {Leib}}\ and\ \bibinfo {author} {\bibfnamefont {M.~J.}\ \bibnamefont
  {Hartmann}},\ }\bibfield  {title} {\enquote {\bibinfo {title} {{Synchronized
  switching in a Josephson junction crystal}},}\ }\href {\doibase
  10.1103/PhysRevLett.112.223603} {\bibfield  {journal} {\bibinfo  {journal}
  {Physical Review Letters}\ }\textbf {\bibinfo {volume} {112}},\ \bibinfo
  {pages} {223603} (\bibinfo {year} {2014})}\BibitemShut {NoStop}%
\bibitem [{\citenamefont {Hartmann}(2016)}]{Hartmann2016}%
  \BibitemOpen
  \bibfield  {author} {\bibinfo {author} {\bibfnamefont {M.~J.}\ \bibnamefont
  {Hartmann}},\ }\bibfield  {title} {\enquote {\bibinfo {title} {{Quantum
  simulation with interacting photons}},}\ }\href {\doibase
  10.1088/2040-8978/18/10/104005} {\bibfield  {journal} {\bibinfo  {journal}
  {Journal of Optics (United Kingdom)}\ }\textbf {\bibinfo {volume} {18}},\
  \bibinfo {pages} {104005} (\bibinfo {year} {2016})}\BibitemShut {NoStop}%
\bibitem [{\citenamefont {Ma}\ \emph {et~al.}(2019)\citenamefont {Ma},
  \citenamefont {Saxberg}, \citenamefont {Owens}, \citenamefont {Leung},
  \citenamefont {Lu}, \citenamefont {Simon},\ and\ \citenamefont
  {Schuster}}]{Ma}%
  \BibitemOpen
  \bibfield  {author} {\bibinfo {author} {\bibfnamefont {R.}~\bibnamefont
  {Ma}}, \bibinfo {author} {\bibfnamefont {B.}~\bibnamefont {Saxberg}},
  \bibinfo {author} {\bibfnamefont {C.}~\bibnamefont {Owens}}, \bibinfo
  {author} {\bibfnamefont {N.}~\bibnamefont {Leung}}, \bibinfo {author}
  {\bibfnamefont {Y.}~\bibnamefont {Lu}}, \bibinfo {author} {\bibfnamefont
  {J.}~\bibnamefont {Simon}}, \ and\ \bibinfo {author} {\bibfnamefont {D.~I.}\
  \bibnamefont {Schuster}},\ }\bibfield  {title} {\enquote {\bibinfo {title}
  {{A dissipatively stabilized Mott insulator of photons}},}\ }\href {\doibase
  10.1038/s41586-019-0897-9} {\bibfield  {journal} {\bibinfo  {journal}
  {Nature}\ }\textbf {\bibinfo {volume} {566}} (\bibinfo {year} {2019}),\
  10.1038/s41586-019-0897-9}\BibitemShut {NoStop}%
\bibitem [{\citenamefont {Chen}\ \emph {et~al.}(2014)\citenamefont {Chen},
  \citenamefont {Megrant}, \citenamefont {Kelly}, \citenamefont {Barends},
  \citenamefont {Bochmann}, \citenamefont {Chen}, \citenamefont {Chiaro},
  \citenamefont {Dunsworth}, \citenamefont {Jeffrey}, \citenamefont {Mutus},
  \citenamefont {O'Malley}, \citenamefont {Neill}, \citenamefont {Roushan},
  \citenamefont {Sank}, \citenamefont {Vainsencher}, \citenamefont {Wenner},
  \citenamefont {White}, \citenamefont {Cleland},\ and\ \citenamefont
  {Martinis}}]{Chen2014a}%
  \BibitemOpen
  \bibfield  {author} {\bibinfo {author} {\bibfnamefont {Z.}~\bibnamefont
  {Chen}}, \bibinfo {author} {\bibfnamefont {A.}~\bibnamefont {Megrant}},
  \bibinfo {author} {\bibfnamefont {J.}~\bibnamefont {Kelly}}, \bibinfo
  {author} {\bibfnamefont {R.}~\bibnamefont {Barends}}, \bibinfo {author}
  {\bibfnamefont {J.}~\bibnamefont {Bochmann}}, \bibinfo {author}
  {\bibfnamefont {Y.}~\bibnamefont {Chen}}, \bibinfo {author} {\bibfnamefont
  {B.}~\bibnamefont {Chiaro}}, \bibinfo {author} {\bibfnamefont
  {A.}~\bibnamefont {Dunsworth}}, \bibinfo {author} {\bibfnamefont
  {E.}~\bibnamefont {Jeffrey}}, \bibinfo {author} {\bibfnamefont {J.~Y.}\
  \bibnamefont {Mutus}}, \bibinfo {author} {\bibfnamefont {P.~J.}\ \bibnamefont
  {O'Malley}}, \bibinfo {author} {\bibfnamefont {C.}~\bibnamefont {Neill}},
  \bibinfo {author} {\bibfnamefont {P.}~\bibnamefont {Roushan}}, \bibinfo
  {author} {\bibfnamefont {D.}~\bibnamefont {Sank}}, \bibinfo {author}
  {\bibfnamefont {A.}~\bibnamefont {Vainsencher}}, \bibinfo {author}
  {\bibfnamefont {J.}~\bibnamefont {Wenner}}, \bibinfo {author} {\bibfnamefont
  {T.~C.}\ \bibnamefont {White}}, \bibinfo {author} {\bibfnamefont {A.~N.}\
  \bibnamefont {Cleland}}, \ and\ \bibinfo {author} {\bibfnamefont {J.~M.}\
  \bibnamefont {Martinis}},\ }\bibfield  {title} {\enquote {\bibinfo {title}
  {{Fabrication and characterization of aluminum airbridges for superconducting
  microwave circuits}},}\ }\href {\doibase 10.1063/1.4863745} {\bibfield
  {journal} {\bibinfo  {journal} {Applied Physics Letters}\ }\textbf {\bibinfo
  {volume} {104}},\ \bibinfo {pages} {052602} (\bibinfo {year}
  {2014})}\BibitemShut {NoStop}%
\bibitem [{\citenamefont {Dunsworth}\ \emph {et~al.}(2018)\citenamefont
  {Dunsworth}, \citenamefont {Barends}, \citenamefont {Chen}, \citenamefont
  {Chen}, \citenamefont {Chiaro}, \citenamefont {Fowler}, \citenamefont
  {Foxen}, \citenamefont {Jeffrey}, \citenamefont {Kelly}, \citenamefont
  {Klimov}, \citenamefont {Lucero}, \citenamefont {Mutus}, \citenamefont
  {Neeley}, \citenamefont {Neill}, \citenamefont {Quintana}, \citenamefont
  {Roushan}, \citenamefont {Sank}, \citenamefont {Vainsencher}, \citenamefont
  {Wenner}, \citenamefont {White}, \citenamefont {Neven}, \citenamefont
  {Martinis},\ and\ \citenamefont {Megrant}}]{Dunsworth2018}%
  \BibitemOpen
  \bibfield  {author} {\bibinfo {author} {\bibfnamefont {A.}~\bibnamefont
  {Dunsworth}}, \bibinfo {author} {\bibfnamefont {R.}~\bibnamefont {Barends}},
  \bibinfo {author} {\bibfnamefont {Y.}~\bibnamefont {Chen}}, \bibinfo {author}
  {\bibfnamefont {Z.}~\bibnamefont {Chen}}, \bibinfo {author} {\bibfnamefont
  {B.}~\bibnamefont {Chiaro}}, \bibinfo {author} {\bibfnamefont
  {A.}~\bibnamefont {Fowler}}, \bibinfo {author} {\bibfnamefont
  {B.}~\bibnamefont {Foxen}}, \bibinfo {author} {\bibfnamefont
  {E.}~\bibnamefont {Jeffrey}}, \bibinfo {author} {\bibfnamefont
  {J.}~\bibnamefont {Kelly}}, \bibinfo {author} {\bibfnamefont {P.~V.}\
  \bibnamefont {Klimov}}, \bibinfo {author} {\bibfnamefont {E.}~\bibnamefont
  {Lucero}}, \bibinfo {author} {\bibfnamefont {J.~Y.}\ \bibnamefont {Mutus}},
  \bibinfo {author} {\bibfnamefont {M.}~\bibnamefont {Neeley}}, \bibinfo
  {author} {\bibfnamefont {C.}~\bibnamefont {Neill}}, \bibinfo {author}
  {\bibfnamefont {C.}~\bibnamefont {Quintana}}, \bibinfo {author}
  {\bibfnamefont {P.}~\bibnamefont {Roushan}}, \bibinfo {author} {\bibfnamefont
  {D.}~\bibnamefont {Sank}}, \bibinfo {author} {\bibfnamefont {A.}~\bibnamefont
  {Vainsencher}}, \bibinfo {author} {\bibfnamefont {J.}~\bibnamefont {Wenner}},
  \bibinfo {author} {\bibfnamefont {T.~C.}\ \bibnamefont {White}}, \bibinfo
  {author} {\bibfnamefont {H.}~\bibnamefont {Neven}}, \bibinfo {author}
  {\bibfnamefont {J.~M.}\ \bibnamefont {Martinis}}, \ and\ \bibinfo {author}
  {\bibfnamefont {A.}~\bibnamefont {Megrant}},\ }\bibfield  {title} {\enquote
  {\bibinfo {title} {{A method for building low loss multi-layer wiring for
  superconducting microwave devices}},}\ }\href {\doibase 10.1063/1.5014033}
  {\bibfield  {journal} {\bibinfo  {journal} {Applied Physics Letters}\
  }\textbf {\bibinfo {volume} {112}},\ \bibinfo {pages} {063502} (\bibinfo
  {year} {2018})}\BibitemShut {NoStop}%
\bibitem [{\citenamefont {Ningyuan}\ \emph {et~al.}(2015)\citenamefont
  {Ningyuan}, \citenamefont {Owens}, \citenamefont {Sommer}, \citenamefont
  {Schuster},\ and\ \citenamefont {Simon}}]{Ningyuan2015}%
  \BibitemOpen
  \bibfield  {author} {\bibinfo {author} {\bibfnamefont {J.}~\bibnamefont
  {Ningyuan}}, \bibinfo {author} {\bibfnamefont {C.}~\bibnamefont {Owens}},
  \bibinfo {author} {\bibfnamefont {A.}~\bibnamefont {Sommer}}, \bibinfo
  {author} {\bibfnamefont {D.}~\bibnamefont {Schuster}}, \ and\ \bibinfo
  {author} {\bibfnamefont {J.}~\bibnamefont {Simon}},\ }\bibfield  {title}
  {\enquote {\bibinfo {title} {{Time- and site-resolved dynamics in a
  topological circuit}},}\ }\href {\doibase 10.1103/PhysRevX.5.021031}
  {\bibfield  {journal} {\bibinfo  {journal} {Physical Review X}\ }\textbf
  {\bibinfo {volume} {5}},\ \bibinfo {pages} {021031} (\bibinfo {year}
  {2015})}\BibitemShut {NoStop}%
\bibitem [{\citenamefont {Lu}\ \emph {et~al.}(2019)\citenamefont {Lu},
  \citenamefont {Jia}, \citenamefont {Su}, \citenamefont {Owens}, \citenamefont
  {Juzeliunas}, \citenamefont {Schuster},\ and\ \citenamefont
  {Simon}}]{Lu2019}%
  \BibitemOpen
  \bibfield  {author} {\bibinfo {author} {\bibfnamefont {Y.}~\bibnamefont
  {Lu}}, \bibinfo {author} {\bibfnamefont {N.}~\bibnamefont {Jia}}, \bibinfo
  {author} {\bibfnamefont {L.}~\bibnamefont {Su}}, \bibinfo {author}
  {\bibfnamefont {C.}~\bibnamefont {Owens}}, \bibinfo {author} {\bibfnamefont
  {G.}~\bibnamefont {Juzeliunas}}, \bibinfo {author} {\bibfnamefont {D.~I.}\
  \bibnamefont {Schuster}}, \ and\ \bibinfo {author} {\bibfnamefont
  {J.}~\bibnamefont {Simon}},\ }\bibfield  {title} {\enquote {\bibinfo {title}
  {{Probing the Berry curvature and Fermi arcs of a Weyl circuit}},}\ }\href
  {\doibase 10.1103/PhysRevB.99.020302} {\bibfield  {journal} {\bibinfo
  {journal} {Physical Review B}\ }\textbf {\bibinfo {volume} {99}},\ \bibinfo
  {pages} {020302(R)} (\bibinfo {year} {2019})}\BibitemShut {NoStop}%
\bibitem [{\citenamefont {Koll{\'{a}}r}, \citenamefont {Fitzpatrick},\ and\
  \citenamefont {Houck}(2019)}]{Kollar2019}%
  \BibitemOpen
  \bibfield  {author} {\bibinfo {author} {\bibfnamefont {A.~J.}\ \bibnamefont
  {Koll{\'{a}}r}}, \bibinfo {author} {\bibfnamefont {M.}~\bibnamefont
  {Fitzpatrick}}, \ and\ \bibinfo {author} {\bibfnamefont {A.~A.}\ \bibnamefont
  {Houck}},\ }\bibfield  {title} {\enquote {\bibinfo {title} {{Hyperbolic
  lattices in circuit quantum electrodynamics}},}\ }\href {\doibase
  10.1038/s41586-019-1348-3} {\bibfield  {journal} {\bibinfo  {journal}
  {Nature}\ }\textbf {\bibinfo {volume} {571}},\ \bibinfo {pages} {45--50}
  (\bibinfo {year} {2019})}\BibitemShut {NoStop}%
\bibitem [{\citenamefont {Cyster}\ \emph {et~al.}(2020)\citenamefont {Cyster},
  \citenamefont {Smith}, \citenamefont {Vaitkus}, \citenamefont {Vogt},
  \citenamefont {Russo},\ and\ \citenamefont {Cole}}]{Cyster2020}%
  \BibitemOpen
  \bibfield  {author} {\bibinfo {author} {\bibfnamefont {M.~J.}\ \bibnamefont
  {Cyster}}, \bibinfo {author} {\bibfnamefont {J.~S.}\ \bibnamefont {Smith}},
  \bibinfo {author} {\bibfnamefont {J.~A.}\ \bibnamefont {Vaitkus}}, \bibinfo
  {author} {\bibfnamefont {N.}~\bibnamefont {Vogt}}, \bibinfo {author}
  {\bibfnamefont {S.~P.}\ \bibnamefont {Russo}}, \ and\ \bibinfo {author}
  {\bibfnamefont {J.~H.}\ \bibnamefont {Cole}},\ }\bibfield  {title} {\enquote
  {\bibinfo {title} {{Effect of atomic structure on the electrical response of
  aluminum oxide tunnel junctions}},}\ }\href {\doibase
  10.1103/physrevresearch.2.013110} {\bibfield  {journal} {\bibinfo  {journal}
  {Physical Review Research}\ }\textbf {\bibinfo {volume} {2}},\ \bibinfo
  {pages} {013110} (\bibinfo {year} {2020})}\BibitemShut {NoStop}%
\bibitem [{\citenamefont {Sarovar}, \citenamefont {Zhang},\ and\ \citenamefont
  {Zeng}(2017)}]{Sarovar2017}%
  \BibitemOpen
  \bibfield  {author} {\bibinfo {author} {\bibfnamefont {M.}~\bibnamefont
  {Sarovar}}, \bibinfo {author} {\bibfnamefont {J.}~\bibnamefont {Zhang}}, \
  and\ \bibinfo {author} {\bibfnamefont {L.}~\bibnamefont {Zeng}},\ }\bibfield
  {title} {\enquote {\bibinfo {title} {{Reliability of analog quantum
  simulation}},}\ }\href {\doibase 10.1140/epjqt/s40507-016-0054-4} {\bibfield
  {journal} {\bibinfo  {journal} {EPJ Quantum Technology}\ }\textbf {\bibinfo
  {volume} {4}} (\bibinfo {year} {2017}),\
  10.1140/epjqt/s40507-016-0054-4}\BibitemShut {NoStop}%
\bibitem [{\citenamefont {Heyl}, \citenamefont {Hauke},\ and\ \citenamefont
  {Zoller}(2019)}]{Heyl2019}%
  \BibitemOpen
  \bibfield  {author} {\bibinfo {author} {\bibfnamefont {M.}~\bibnamefont
  {Heyl}}, \bibinfo {author} {\bibfnamefont {P.}~\bibnamefont {Hauke}}, \ and\
  \bibinfo {author} {\bibfnamefont {P.}~\bibnamefont {Zoller}},\ }\bibfield
  {title} {\enquote {\bibinfo {title} {{Quantum localization bounds Trotter
  errors in digital quantum simulation}},}\ }\href {\doibase
  10.1126/sciadv.aau8342} {\bibfield  {journal} {\bibinfo  {journal} {Science
  Advances}\ }\textbf {\bibinfo {volume} {5}},\ \bibinfo {pages} {eaau8342}
  (\bibinfo {year} {2019})},\ \Eprint {http://arxiv.org/abs/1806.11123}
  {arXiv:1806.11123} \BibitemShut {NoStop}%
\bibitem [{\citenamefont {Derbyshire}\ \emph {et~al.}(2020)\citenamefont
  {Derbyshire}, \citenamefont {Malo}, \citenamefont {Daley}, \citenamefont
  {Kashefi},\ and\ \citenamefont {Wallden}}]{Derbyshire2020}%
  \BibitemOpen
  \bibfield  {author} {\bibinfo {author} {\bibfnamefont {E.}~\bibnamefont
  {Derbyshire}}, \bibinfo {author} {\bibfnamefont {J.~Y.}\ \bibnamefont
  {Malo}}, \bibinfo {author} {\bibfnamefont {A.~J.}\ \bibnamefont {Daley}},
  \bibinfo {author} {\bibfnamefont {E.}~\bibnamefont {Kashefi}}, \ and\
  \bibinfo {author} {\bibfnamefont {P.}~\bibnamefont {Wallden}},\ }\bibfield
  {title} {\enquote {\bibinfo {title} {{Randomized benchmarking in the analogue
  setting}},}\ }\href@noop {} {\bibfield  {journal} {\bibinfo  {journal}
  {Quantum Science and Technology}\ }\textbf {\bibinfo {volume} {5}},\ \bibinfo
  {pages} {034001} (\bibinfo {year} {2020})}\BibitemShut {NoStop}%
\bibitem [{\citenamefont {Arute}\ \emph {et~al.}(2019)\citenamefont {Arute},
  \citenamefont {Arya}, \citenamefont {Babbush}, \citenamefont {Bacon},
  \citenamefont {Bardin}, \citenamefont {Barends}, \citenamefont {Biswas},
  \citenamefont {Boixo}, \citenamefont {Brandao}, \citenamefont {Buell},
  \citenamefont {Burkett}, \citenamefont {Chen}, \citenamefont {Chen},
  \citenamefont {Chiaro}, \citenamefont {Collins}, \citenamefont {Courtney},
  \citenamefont {Dunsworth}, \citenamefont {Farhi}, \citenamefont {Foxen},
  \citenamefont {Fowler}, \citenamefont {Gidney}, \citenamefont {Giustina},
  \citenamefont {Graff}, \citenamefont {Guerin}, \citenamefont {Habegger},
  \citenamefont {Harrigan}, \citenamefont {Hartmann}, \citenamefont {Ho},
  \citenamefont {Hoffmann}, \citenamefont {Huang}, \citenamefont {Humble},
  \citenamefont {Isakov}, \citenamefont {Jeffrey}, \citenamefont {Jiang},
  \citenamefont {Kafri}, \citenamefont {Kechedzhi}, \citenamefont {Kelly},
  \citenamefont {Klimov}, \citenamefont {Knysh}, \citenamefont {Korotkov},
  \citenamefont {Kostritsa}, \citenamefont {Landhuis}, \citenamefont
  {Lindmark}, \citenamefont {Lucero}, \citenamefont {Lyakh}, \citenamefont
  {Mandra}, \citenamefont {McClean}, \citenamefont {McEwen}, \citenamefont
  {Megrant}, \citenamefont {Mi}, \citenamefont {Michielsen}, \citenamefont
  {Mohseni}, \citenamefont {Mutus}, \citenamefont {Naaman}, \citenamefont
  {Neeley}, \citenamefont {Neill}, \citenamefont {Niu}, \citenamefont {Ostby},
  \citenamefont {Petukhov}, \citenamefont {Platt}, \citenamefont {Quintana},
  \citenamefont {Rieffel}, \citenamefont {Roushan}, \citenamefont {Rubin},
  \citenamefont {Sank}, \citenamefont {Satzinger}, \citenamefont {Smelyanskiy},
  \citenamefont {Sung}, \citenamefont {Trevithick}, \citenamefont
  {Vainsencher}, \citenamefont {Villalonga}, \citenamefont {White},
  \citenamefont {Yao}, \citenamefont {Yeh}, \citenamefont {Zalcman},
  \citenamefont {Neven},\ and\ \citenamefont {Martinis}}]{Arute2019}%
  \BibitemOpen
  \bibfield  {author} {\bibinfo {author} {\bibfnamefont {F.}~\bibnamefont
  {Arute}}, \bibinfo {author} {\bibfnamefont {K.}~\bibnamefont {Arya}},
  \bibinfo {author} {\bibfnamefont {R.}~\bibnamefont {Babbush}}, \bibinfo
  {author} {\bibfnamefont {D.}~\bibnamefont {Bacon}}, \bibinfo {author}
  {\bibfnamefont {J.~C.}\ \bibnamefont {Bardin}}, \bibinfo {author}
  {\bibfnamefont {R.}~\bibnamefont {Barends}}, \bibinfo {author} {\bibfnamefont
  {R.}~\bibnamefont {Biswas}}, \bibinfo {author} {\bibfnamefont
  {S.}~\bibnamefont {Boixo}}, \bibinfo {author} {\bibfnamefont {F.~G. S.~L.}\
  \bibnamefont {Brandao}}, \bibinfo {author} {\bibfnamefont {D.~A.}\
  \bibnamefont {Buell}}, \bibinfo {author} {\bibfnamefont {B.}~\bibnamefont
  {Burkett}}, \bibinfo {author} {\bibfnamefont {Y.}~\bibnamefont {Chen}},
  \bibinfo {author} {\bibfnamefont {Z.}~\bibnamefont {Chen}}, \bibinfo {author}
  {\bibfnamefont {B.}~\bibnamefont {Chiaro}}, \bibinfo {author} {\bibfnamefont
  {R.}~\bibnamefont {Collins}}, \bibinfo {author} {\bibfnamefont
  {W.}~\bibnamefont {Courtney}}, \bibinfo {author} {\bibfnamefont
  {A.}~\bibnamefont {Dunsworth}}, \bibinfo {author} {\bibfnamefont
  {E.}~\bibnamefont {Farhi}}, \bibinfo {author} {\bibfnamefont
  {B.}~\bibnamefont {Foxen}}, \bibinfo {author} {\bibfnamefont
  {A.}~\bibnamefont {Fowler}}, \bibinfo {author} {\bibfnamefont
  {C.}~\bibnamefont {Gidney}}, \bibinfo {author} {\bibfnamefont
  {M.}~\bibnamefont {Giustina}}, \bibinfo {author} {\bibfnamefont
  {R.}~\bibnamefont {Graff}}, \bibinfo {author} {\bibfnamefont
  {K.}~\bibnamefont {Guerin}}, \bibinfo {author} {\bibfnamefont
  {S.}~\bibnamefont {Habegger}}, \bibinfo {author} {\bibfnamefont {M.~P.}\
  \bibnamefont {Harrigan}}, \bibinfo {author} {\bibfnamefont {M.~J.}\
  \bibnamefont {Hartmann}}, \bibinfo {author} {\bibfnamefont {A.}~\bibnamefont
  {Ho}}, \bibinfo {author} {\bibfnamefont {M.~R.}\ \bibnamefont {Hoffmann}},
  \bibinfo {author} {\bibfnamefont {T.}~\bibnamefont {Huang}}, \bibinfo
  {author} {\bibfnamefont {T.~S.}\ \bibnamefont {Humble}}, \bibinfo {author}
  {\bibfnamefont {S.~V.}\ \bibnamefont {Isakov}}, \bibinfo {author}
  {\bibfnamefont {E.}~\bibnamefont {Jeffrey}}, \bibinfo {author} {\bibfnamefont
  {Z.}~\bibnamefont {Jiang}}, \bibinfo {author} {\bibfnamefont
  {D.}~\bibnamefont {Kafri}}, \bibinfo {author} {\bibfnamefont
  {K.}~\bibnamefont {Kechedzhi}}, \bibinfo {author} {\bibfnamefont
  {J.}~\bibnamefont {Kelly}}, \bibinfo {author} {\bibfnamefont {P.~V.}\
  \bibnamefont {Klimov}}, \bibinfo {author} {\bibfnamefont {S.}~\bibnamefont
  {Knysh}}, \bibinfo {author} {\bibfnamefont {A.~N.}\ \bibnamefont {Korotkov}},
  \bibinfo {author} {\bibfnamefont {F.}~\bibnamefont {Kostritsa}}, \bibinfo
  {author} {\bibfnamefont {D.}~\bibnamefont {Landhuis}}, \bibinfo {author}
  {\bibfnamefont {M.}~\bibnamefont {Lindmark}}, \bibinfo {author}
  {\bibfnamefont {E.}~\bibnamefont {Lucero}}, \bibinfo {author} {\bibfnamefont
  {D.}~\bibnamefont {Lyakh}}, \bibinfo {author} {\bibfnamefont
  {S.}~\bibnamefont {Mandra}}, \bibinfo {author} {\bibfnamefont {J.~R.}\
  \bibnamefont {McClean}}, \bibinfo {author} {\bibfnamefont {M.}~\bibnamefont
  {McEwen}}, \bibinfo {author} {\bibfnamefont {A.}~\bibnamefont {Megrant}},
  \bibinfo {author} {\bibfnamefont {X.}~\bibnamefont {Mi}}, \bibinfo {author}
  {\bibfnamefont {K.}~\bibnamefont {Michielsen}}, \bibinfo {author}
  {\bibfnamefont {M.}~\bibnamefont {Mohseni}}, \bibinfo {author} {\bibfnamefont
  {J.}~\bibnamefont {Mutus}}, \bibinfo {author} {\bibfnamefont
  {O.}~\bibnamefont {Naaman}}, \bibinfo {author} {\bibfnamefont
  {M.}~\bibnamefont {Neeley}}, \bibinfo {author} {\bibfnamefont
  {C.}~\bibnamefont {Neill}}, \bibinfo {author} {\bibfnamefont {M.~Y.}\
  \bibnamefont {Niu}}, \bibinfo {author} {\bibfnamefont {E.}~\bibnamefont
  {Ostby}}, \bibinfo {author} {\bibfnamefont {A.}~\bibnamefont {Petukhov}},
  \bibinfo {author} {\bibfnamefont {J.~C.}\ \bibnamefont {Platt}}, \bibinfo
  {author} {\bibfnamefont {C.}~\bibnamefont {Quintana}}, \bibinfo {author}
  {\bibfnamefont {E.~G.}\ \bibnamefont {Rieffel}}, \bibinfo {author}
  {\bibfnamefont {P.}~\bibnamefont {Roushan}}, \bibinfo {author} {\bibfnamefont
  {N.~C.}\ \bibnamefont {Rubin}}, \bibinfo {author} {\bibfnamefont
  {D.}~\bibnamefont {Sank}}, \bibinfo {author} {\bibfnamefont {K.~J.}\
  \bibnamefont {Satzinger}}, \bibinfo {author} {\bibfnamefont {V.}~\bibnamefont
  {Smelyanskiy}}, \bibinfo {author} {\bibfnamefont {K.~J.}\ \bibnamefont
  {Sung}}, \bibinfo {author} {\bibfnamefont {M.~D.}\ \bibnamefont
  {Trevithick}}, \bibinfo {author} {\bibfnamefont {A.}~\bibnamefont
  {Vainsencher}}, \bibinfo {author} {\bibfnamefont {B.}~\bibnamefont
  {Villalonga}}, \bibinfo {author} {\bibfnamefont {T.}~\bibnamefont {White}},
  \bibinfo {author} {\bibfnamefont {Z.~J.}\ \bibnamefont {Yao}}, \bibinfo
  {author} {\bibfnamefont {P.}~\bibnamefont {Yeh}}, \bibinfo {author}
  {\bibfnamefont {A.}~\bibnamefont {Zalcman}}, \bibinfo {author} {\bibfnamefont
  {H.}~\bibnamefont {Neven}}, \ and\ \bibinfo {author} {\bibfnamefont {J.~M.}\
  \bibnamefont {Martinis}},\ }\bibfield  {title} {\enquote {\bibinfo {title}
  {{Quantum supremacy using a programmable superconducting processor}},}\
  }\href {\doibase 10.1038/s41586-019-1666-5} {\bibfield  {journal} {\bibinfo
  {journal} {Nature}\ }\textbf {\bibinfo {volume} {574}},\ \bibinfo {pages}
  {505} (\bibinfo {year} {2019})}\BibitemShut {NoStop}%
\bibitem [{\citenamefont {Salath{\'{e}}}\ \emph {et~al.}(2015)\citenamefont
  {Salath{\'{e}}}, \citenamefont {Mondal}, \citenamefont {Oppliger},
  \citenamefont {Heinsoo}, \citenamefont {Kurpiers}, \citenamefont {Poto},
  \citenamefont {Mezzacapo}, \citenamefont {Heras}, \citenamefont {Lamata},
  \citenamefont {Solano}, \citenamefont {Filipp},\ and\ \citenamefont
  {Wallraff}}]{Salathe2015}%
  \BibitemOpen
  \bibfield  {author} {\bibinfo {author} {\bibfnamefont {Y.}~\bibnamefont
  {Salath{\'{e}}}}, \bibinfo {author} {\bibfnamefont {M.}~\bibnamefont
  {Mondal}}, \bibinfo {author} {\bibfnamefont {M.}~\bibnamefont {Oppliger}},
  \bibinfo {author} {\bibfnamefont {J.}~\bibnamefont {Heinsoo}}, \bibinfo
  {author} {\bibfnamefont {P.}~\bibnamefont {Kurpiers}}, \bibinfo {author}
  {\bibfnamefont {A.}~\bibnamefont {Poto}}, \bibinfo {author} {\bibfnamefont
  {A.}~\bibnamefont {Mezzacapo}}, \bibinfo {author} {\bibfnamefont {U.~L.}\
  \bibnamefont {Heras}}, \bibinfo {author} {\bibfnamefont {L.}~\bibnamefont
  {Lamata}}, \bibinfo {author} {\bibfnamefont {E.}~\bibnamefont {Solano}},
  \bibinfo {author} {\bibfnamefont {S.}~\bibnamefont {Filipp}}, \ and\ \bibinfo
  {author} {\bibfnamefont {A.}~\bibnamefont {Wallraff}},\ }\bibfield  {title}
  {\enquote {\bibinfo {title} {{Digital Quantum Simulation of Spin Models with
  Circuit Quantum Electrodynamics}},}\ }\href {\doibase
  10.1103/PhysRevX.5.021027} {\bibfield  {journal} {\bibinfo  {journal}
  {Physical Review X}\ }\textbf {\bibinfo {volume} {5}},\ \bibinfo {pages}
  {021027} (\bibinfo {year} {2015})}\BibitemShut {NoStop}%
\bibitem [{\citenamefont {Heras}\ \emph {et~al.}(2014)\citenamefont {Heras},
  \citenamefont {Mezzacapo}, \citenamefont {Lamata}, \citenamefont {Filipp},
  \citenamefont {Wallraff},\ and\ \citenamefont {Solano}}]{Heras2014}%
  \BibitemOpen
  \bibfield  {author} {\bibinfo {author} {\bibfnamefont {U.~L.}\ \bibnamefont
  {Heras}}, \bibinfo {author} {\bibfnamefont {A.}~\bibnamefont {Mezzacapo}},
  \bibinfo {author} {\bibfnamefont {L.}~\bibnamefont {Lamata}}, \bibinfo
  {author} {\bibfnamefont {S.}~\bibnamefont {Filipp}}, \bibinfo {author}
  {\bibfnamefont {A.}~\bibnamefont {Wallraff}}, \ and\ \bibinfo {author}
  {\bibfnamefont {E.}~\bibnamefont {Solano}},\ }\bibfield  {title} {\enquote
  {\bibinfo {title} {{Digital quantum simulation of spin systems in
  superconducting circuits}},}\ }\href {\doibase
  10.1103/PhysRevLett.112.200501} {\bibfield  {journal} {\bibinfo  {journal}
  {Physical Review Letters}\ }\textbf {\bibinfo {volume} {112}},\ \bibinfo
  {pages} {200501} (\bibinfo {year} {2014})}\BibitemShut {NoStop}%
\bibitem [{\citenamefont {Barends}\ \emph {et~al.}(2015)\citenamefont
  {Barends}, \citenamefont {Lamata}, \citenamefont {Kelly}, \citenamefont
  {Garc{\'{i}}a-{\'{A}}lvarez}, \citenamefont {Fowler}, \citenamefont
  {Megrant}, \citenamefont {Jeffrey}, \citenamefont {White}, \citenamefont
  {Sank}, \citenamefont {Mutus}, \citenamefont {Campbell}, \citenamefont
  {Chen}, \citenamefont {Chen}, \citenamefont {Chiaro}, \citenamefont
  {Dunsworth}, \citenamefont {Hoi}, \citenamefont {Neill}, \citenamefont
  {O'Malley}, \citenamefont {Quintana}, \citenamefont {Roushan}, \citenamefont
  {Vainsencher}, \citenamefont {Wenner}, \citenamefont {Solano},\ and\
  \citenamefont {Martinis}}]{Barends2015}%
  \BibitemOpen
  \bibfield  {author} {\bibinfo {author} {\bibfnamefont {R.}~\bibnamefont
  {Barends}}, \bibinfo {author} {\bibfnamefont {L.}~\bibnamefont {Lamata}},
  \bibinfo {author} {\bibfnamefont {J.}~\bibnamefont {Kelly}}, \bibinfo
  {author} {\bibfnamefont {L.}~\bibnamefont {Garc{\'{i}}a-{\'{A}}lvarez}},
  \bibinfo {author} {\bibfnamefont {A.~G.}\ \bibnamefont {Fowler}}, \bibinfo
  {author} {\bibfnamefont {A.}~\bibnamefont {Megrant}}, \bibinfo {author}
  {\bibfnamefont {E.}~\bibnamefont {Jeffrey}}, \bibinfo {author} {\bibfnamefont
  {T.~C.}\ \bibnamefont {White}}, \bibinfo {author} {\bibfnamefont
  {D.}~\bibnamefont {Sank}}, \bibinfo {author} {\bibfnamefont {J.~Y.}\
  \bibnamefont {Mutus}}, \bibinfo {author} {\bibfnamefont {B.}~\bibnamefont
  {Campbell}}, \bibinfo {author} {\bibfnamefont {Y.}~\bibnamefont {Chen}},
  \bibinfo {author} {\bibfnamefont {Z.}~\bibnamefont {Chen}}, \bibinfo {author}
  {\bibfnamefont {B.}~\bibnamefont {Chiaro}}, \bibinfo {author} {\bibfnamefont
  {A.}~\bibnamefont {Dunsworth}}, \bibinfo {author} {\bibfnamefont {I.~C.}\
  \bibnamefont {Hoi}}, \bibinfo {author} {\bibfnamefont {C.}~\bibnamefont
  {Neill}}, \bibinfo {author} {\bibfnamefont {P.~J.}\ \bibnamefont {O'Malley}},
  \bibinfo {author} {\bibfnamefont {C.}~\bibnamefont {Quintana}}, \bibinfo
  {author} {\bibfnamefont {P.}~\bibnamefont {Roushan}}, \bibinfo {author}
  {\bibfnamefont {A.}~\bibnamefont {Vainsencher}}, \bibinfo {author}
  {\bibfnamefont {J.}~\bibnamefont {Wenner}}, \bibinfo {author} {\bibfnamefont
  {E.}~\bibnamefont {Solano}}, \ and\ \bibinfo {author} {\bibfnamefont {J.~M.}\
  \bibnamefont {Martinis}},\ }\bibfield  {title} {\enquote {\bibinfo {title}
  {{Digital quantum simulation of fermionic models with a superconducting
  circuit}},}\ }\href {\doibase 10.1038/ncomms8654} {\bibfield  {journal}
  {\bibinfo  {journal} {Nature Communications}\ }\textbf {\bibinfo {volume}
  {6}},\ \bibinfo {pages} {8654} (\bibinfo {year} {2015})}\BibitemShut
  {NoStop}%
\bibitem [{\citenamefont {Heras}\ \emph {et~al.}(2015)\citenamefont {Heras},
  \citenamefont {Garc{\'{i}}a-{\'{a}}lvarez}, \citenamefont {Mezzacapo},
  \citenamefont {Solano},\ and\ \citenamefont {Lamata}}]{Heras2015}%
  \BibitemOpen
  \bibfield  {author} {\bibinfo {author} {\bibfnamefont {U.~L.}\ \bibnamefont
  {Heras}}, \bibinfo {author} {\bibfnamefont {L.}~\bibnamefont
  {Garc{\'{i}}a-{\'{a}}lvarez}}, \bibinfo {author} {\bibfnamefont
  {A.}~\bibnamefont {Mezzacapo}}, \bibinfo {author} {\bibfnamefont
  {E.}~\bibnamefont {Solano}}, \ and\ \bibinfo {author} {\bibfnamefont
  {L.}~\bibnamefont {Lamata}},\ }\bibfield  {title} {\enquote {\bibinfo {title}
  {{Fermionic models with superconducting circuits}},}\ }\href {\doibase
  10.1140/epjqt/s40507-015-0021-5} {\bibfield  {journal} {\bibinfo  {journal}
  {EPJ Quantum Technol.}\ }\textbf {\bibinfo {volume} {2}} (\bibinfo {year}
  {2015}),\ 10.1140/epjqt/s40507-015-0021-5}\BibitemShut {NoStop}%
\bibitem [{\citenamefont {Mezzacapo}\ \emph {et~al.}(2014)\citenamefont
  {Mezzacapo}, \citenamefont {{Las Heras}}, \citenamefont {Pedernales},
  \citenamefont {DiCarlo}, \citenamefont {Solano},\ and\ \citenamefont
  {Lamata}}]{Mezzacapo2014}%
  \BibitemOpen
  \bibfield  {author} {\bibinfo {author} {\bibfnamefont {A.}~\bibnamefont
  {Mezzacapo}}, \bibinfo {author} {\bibfnamefont {U.}~\bibnamefont {{Las
  Heras}}}, \bibinfo {author} {\bibfnamefont {J.~S.}\ \bibnamefont
  {Pedernales}}, \bibinfo {author} {\bibfnamefont {L.}~\bibnamefont {DiCarlo}},
  \bibinfo {author} {\bibfnamefont {E.}~\bibnamefont {Solano}}, \ and\ \bibinfo
  {author} {\bibfnamefont {L.}~\bibnamefont {Lamata}},\ }\bibfield  {title}
  {\enquote {\bibinfo {title} {{Digital quantum rabi and dicke models in
  superconducting circuits}},}\ }\href {\doibase 10.1038/srep07482} {\bibfield
  {journal} {\bibinfo  {journal} {Scientific Reports}\ }\textbf {\bibinfo
  {volume} {4}} (\bibinfo {year} {2014}),\ 10.1038/srep07482}\BibitemShut
  {NoStop}%
\bibitem [{\citenamefont {McClean}\ \emph {et~al.}(2016)\citenamefont
  {McClean}, \citenamefont {Romero}, \citenamefont {Babbush},\ and\
  \citenamefont {Aspuru-Guzik}}]{McClean2016}%
  \BibitemOpen
  \bibfield  {author} {\bibinfo {author} {\bibfnamefont {J.~R.}\ \bibnamefont
  {McClean}}, \bibinfo {author} {\bibfnamefont {J.}~\bibnamefont {Romero}},
  \bibinfo {author} {\bibfnamefont {R.}~\bibnamefont {Babbush}}, \ and\
  \bibinfo {author} {\bibfnamefont {A.}~\bibnamefont {Aspuru-Guzik}},\
  }\bibfield  {title} {\enquote {\bibinfo {title} {{The theory of variational
  hybrid quantum-classical algorithms The theory of variational hybrid
  quantum-classical algorithms}},}\ }\href@noop {} {\bibfield  {journal}
  {\bibinfo  {journal} {New Journal of Physics}\ }\textbf {\bibinfo {volume}
  {18}},\ \bibinfo {pages} {023023} (\bibinfo {year} {2016})}\BibitemShut
  {NoStop}%
\bibitem [{\citenamefont {Shor}(1994)}]{Shor2002}%
  \BibitemOpen
  \bibfield  {author} {\bibinfo {author} {\bibfnamefont {P.}~\bibnamefont
  {Shor}},\ }\bibfield  {title} {\enquote {\bibinfo {title} {{Algorithms for
  quantum computation: discrete logarithms and factoring}},}\ }in\ \href
  {\doibase 10.1109/sfcs.1994.365700} {\emph {\bibinfo {booktitle} {Proceedings
  35th Annual Symposium on Foundations of Computer Science}}}\ (\bibinfo
  {publisher} {IEEE Comput. Soc. Press},\ \bibinfo {year} {1994})\ pp.\
  \bibinfo {pages} {124--134}\BibitemShut {NoStop}%
\bibitem [{\citenamefont {Harrow}, \citenamefont {Hassidim},\ and\
  \citenamefont {Lloyd}(2009)}]{Harrow2009}%
  \BibitemOpen
  \bibfield  {author} {\bibinfo {author} {\bibfnamefont {A.~W.}\ \bibnamefont
  {Harrow}}, \bibinfo {author} {\bibfnamefont {A.}~\bibnamefont {Hassidim}}, \
  and\ \bibinfo {author} {\bibfnamefont {S.}~\bibnamefont {Lloyd}},\ }\bibfield
   {title} {\enquote {\bibinfo {title} {{Quantum algorithm for linear systems
  of equations}},}\ }\href {\doibase 10.1103/PhysRevLett.103.150502} {\bibfield
   {journal} {\bibinfo  {journal} {Physical Review Letters}\ }\textbf {\bibinfo
  {volume} {103}},\ \bibinfo {pages} {150502} (\bibinfo {year}
  {2009})}\BibitemShut {NoStop}%
\bibitem [{\citenamefont {Blatt}\ and\ \citenamefont {Roos}(2012)}]{Blatt2012}%
  \BibitemOpen
  \bibfield  {author} {\bibinfo {author} {\bibfnamefont {R.}~\bibnamefont
  {Blatt}}\ and\ \bibinfo {author} {\bibfnamefont {C.~F.}\ \bibnamefont
  {Roos}},\ }\bibfield  {title} {\enquote {\bibinfo {title} {{Quantum
  simulations with trapped ions}},}\ }\href {\doibase 10.1038/nphys2252}
  {\bibfield  {journal} {\bibinfo  {journal} {Nature Physics}\ }\textbf
  {\bibinfo {volume} {8}} (\bibinfo {year} {2012}),\
  10.1038/nphys2252}\BibitemShut {NoStop}%
\bibitem [{\citenamefont {Tranter}\ \emph {et~al.}(2019)\citenamefont
  {Tranter}, \citenamefont {Love}, \citenamefont {Mintert}, \citenamefont
  {Wiebe},\ and\ \citenamefont {Coveney}}]{Tranter2019}%
  \BibitemOpen
  \bibfield  {author} {\bibinfo {author} {\bibfnamefont {A.}~\bibnamefont
  {Tranter}}, \bibinfo {author} {\bibfnamefont {P.~J.}\ \bibnamefont {Love}},
  \bibinfo {author} {\bibfnamefont {F.}~\bibnamefont {Mintert}}, \bibinfo
  {author} {\bibfnamefont {N.}~\bibnamefont {Wiebe}}, \ and\ \bibinfo {author}
  {\bibfnamefont {P.~V.}\ \bibnamefont {Coveney}},\ }\bibfield  {title}
  {\enquote {\bibinfo {title} {{Ordering of Trotterization: Impact on Errors in
  Quantum Simulation of Electronic Structure}},}\ }\href {\doibase
  10.3390/e21121218} {\bibfield  {journal} {\bibinfo  {journal} {Entropy}\
  }\textbf {\bibinfo {volume} {21}},\ \bibinfo {pages} {1218} (\bibinfo {year}
  {2019})}\BibitemShut {NoStop}%
\bibitem [{\citenamefont {Childs}\ \emph {et~al.}(2019)\citenamefont {Childs},
  \citenamefont {Su}, \citenamefont {Tran}, \citenamefont {Wiebe},\ and\
  \citenamefont {Zhu}}]{Childs2019}%
  \BibitemOpen
  \bibfield  {author} {\bibinfo {author} {\bibfnamefont {A.~M.}\ \bibnamefont
  {Childs}}, \bibinfo {author} {\bibfnamefont {Y.}~\bibnamefont {Su}}, \bibinfo
  {author} {\bibfnamefont {M.~C.}\ \bibnamefont {Tran}}, \bibinfo {author}
  {\bibfnamefont {N.}~\bibnamefont {Wiebe}}, \ and\ \bibinfo {author}
  {\bibfnamefont {S.}~\bibnamefont {Zhu}},\ }\bibfield  {title} {\enquote
  {\bibinfo {title} {{A Theory of Trotter Error}},}\ }\href
  {http://arxiv.org/abs/1912.08854} {\  (\bibinfo {year} {2019})},\ \Eprint
  {http://arxiv.org/abs/1912.08854} {arXiv:1912.08854} \BibitemShut {NoStop}%
\bibitem [{\citenamefont {Preskill}(2018)}]{Preskill2018}%
  \BibitemOpen
  \bibfield  {author} {\bibinfo {author} {\bibfnamefont {J.}~\bibnamefont
  {Preskill}},\ }\bibfield  {title} {\enquote {\bibinfo {title} {{Quantum
  Computing in the NISQ era and beyond}},}\ }\href {\doibase
  10.22331/q-2018-08-06-79} {\bibfield  {journal} {\bibinfo  {journal}
  {Quantum}\ }\textbf {\bibinfo {volume} {2}},\ \bibinfo {pages} {79} (\bibinfo
  {year} {2018})}\BibitemShut {NoStop}%
\bibitem [{\citenamefont {O'Malley}\ \emph {et~al.}(2016)\citenamefont
  {O'Malley}, \citenamefont {Babbush}, \citenamefont {Kivlichan}, \citenamefont
  {Romero}, \citenamefont {McClean}, \citenamefont {Barends}, \citenamefont
  {Kelly}, \citenamefont {Roushan}, \citenamefont {Tranter}, \citenamefont
  {Ding}, \citenamefont {Campbell}, \citenamefont {Chen}, \citenamefont {Chen},
  \citenamefont {Chiaro}, \citenamefont {Dunsworth}, \citenamefont {Fowler},
  \citenamefont {Jeffrey}, \citenamefont {Lucero}, \citenamefont {Megrant},
  \citenamefont {Mutus}, \citenamefont {Neeley}, \citenamefont {Neill},
  \citenamefont {Quintana}, \citenamefont {Sank}, \citenamefont {Vainsencher},
  \citenamefont {Wenner}, \citenamefont {White}, \citenamefont {Coveney},
  \citenamefont {Love}, \citenamefont {Neven}, \citenamefont {Aspuru-Guzik},\
  and\ \citenamefont {Martinis}}]{OMalley2016}%
  \BibitemOpen
  \bibfield  {author} {\bibinfo {author} {\bibfnamefont {P.~J.}\ \bibnamefont
  {O'Malley}}, \bibinfo {author} {\bibfnamefont {R.}~\bibnamefont {Babbush}},
  \bibinfo {author} {\bibfnamefont {I.~D.}\ \bibnamefont {Kivlichan}}, \bibinfo
  {author} {\bibfnamefont {J.}~\bibnamefont {Romero}}, \bibinfo {author}
  {\bibfnamefont {J.~R.}\ \bibnamefont {McClean}}, \bibinfo {author}
  {\bibfnamefont {R.}~\bibnamefont {Barends}}, \bibinfo {author} {\bibfnamefont
  {J.}~\bibnamefont {Kelly}}, \bibinfo {author} {\bibfnamefont
  {P.}~\bibnamefont {Roushan}}, \bibinfo {author} {\bibfnamefont
  {A.}~\bibnamefont {Tranter}}, \bibinfo {author} {\bibfnamefont
  {N.}~\bibnamefont {Ding}}, \bibinfo {author} {\bibfnamefont {B.}~\bibnamefont
  {Campbell}}, \bibinfo {author} {\bibfnamefont {Y.}~\bibnamefont {Chen}},
  \bibinfo {author} {\bibfnamefont {Z.}~\bibnamefont {Chen}}, \bibinfo {author}
  {\bibfnamefont {B.}~\bibnamefont {Chiaro}}, \bibinfo {author} {\bibfnamefont
  {A.}~\bibnamefont {Dunsworth}}, \bibinfo {author} {\bibfnamefont {A.~G.}\
  \bibnamefont {Fowler}}, \bibinfo {author} {\bibfnamefont {E.}~\bibnamefont
  {Jeffrey}}, \bibinfo {author} {\bibfnamefont {E.}~\bibnamefont {Lucero}},
  \bibinfo {author} {\bibfnamefont {A.}~\bibnamefont {Megrant}}, \bibinfo
  {author} {\bibfnamefont {J.~Y.}\ \bibnamefont {Mutus}}, \bibinfo {author}
  {\bibfnamefont {M.}~\bibnamefont {Neeley}}, \bibinfo {author} {\bibfnamefont
  {C.}~\bibnamefont {Neill}}, \bibinfo {author} {\bibfnamefont
  {C.}~\bibnamefont {Quintana}}, \bibinfo {author} {\bibfnamefont
  {D.}~\bibnamefont {Sank}}, \bibinfo {author} {\bibfnamefont {A.}~\bibnamefont
  {Vainsencher}}, \bibinfo {author} {\bibfnamefont {J.}~\bibnamefont {Wenner}},
  \bibinfo {author} {\bibfnamefont {T.~C.}\ \bibnamefont {White}}, \bibinfo
  {author} {\bibfnamefont {P.~V.}\ \bibnamefont {Coveney}}, \bibinfo {author}
  {\bibfnamefont {P.~J.}\ \bibnamefont {Love}}, \bibinfo {author}
  {\bibfnamefont {H.}~\bibnamefont {Neven}}, \bibinfo {author} {\bibfnamefont
  {A.}~\bibnamefont {Aspuru-Guzik}}, \ and\ \bibinfo {author} {\bibfnamefont
  {J.~M.}\ \bibnamefont {Martinis}},\ }\bibfield  {title} {\enquote {\bibinfo
  {title} {{Scalable quantum simulation of molecular energies}},}\ }\href
  {\doibase 10.1103/PhysRevX.6.031007} {\bibfield  {journal} {\bibinfo
  {journal} {Physical Review X}\ }\textbf {\bibinfo {volume} {6}},\ \bibinfo
  {pages} {031007} (\bibinfo {year} {2016})}\BibitemShut {NoStop}%
\bibitem [{\citenamefont {Kandala}\ \emph {et~al.}(2017)\citenamefont
  {Kandala}, \citenamefont {Mezzacapo}, \citenamefont {Temme}, \citenamefont
  {Takita}, \citenamefont {Brink}, \citenamefont {Chow},\ and\ \citenamefont
  {Gambetta}}]{Kandala:2017fa}%
  \BibitemOpen
  \bibfield  {author} {\bibinfo {author} {\bibfnamefont {A.}~\bibnamefont
  {Kandala}}, \bibinfo {author} {\bibfnamefont {A.}~\bibnamefont {Mezzacapo}},
  \bibinfo {author} {\bibfnamefont {K.}~\bibnamefont {Temme}}, \bibinfo
  {author} {\bibfnamefont {M.}~\bibnamefont {Takita}}, \bibinfo {author}
  {\bibfnamefont {M.}~\bibnamefont {Brink}}, \bibinfo {author} {\bibfnamefont
  {J.~M.}\ \bibnamefont {Chow}}, \ and\ \bibinfo {author} {\bibfnamefont
  {J.~M.}\ \bibnamefont {Gambetta}},\ }\bibfield  {title} {\enquote {\bibinfo
  {title} {{Hardware-efficient variational quantum eigensolver for small
  molecules and quantum magnets}},}\ }\href {\doibase 10.1038/nature23879}
  {\bibfield  {journal} {\bibinfo  {journal} {Nature}\ }\textbf {\bibinfo
  {volume} {549}},\ \bibinfo {pages} {242} (\bibinfo {year}
  {2017})}\BibitemShut {NoStop}%
\bibitem [{\citenamefont {Kokail}\ \emph {et~al.}(2019)\citenamefont {Kokail},
  \citenamefont {Maier}, \citenamefont {Bijnen}, \citenamefont {Brydges},\ and\
  \citenamefont {Joshi}}]{Kokail:2019fw}%
  \BibitemOpen
  \bibfield  {author} {\bibinfo {author} {\bibfnamefont {C.}~\bibnamefont
  {Kokail}}, \bibinfo {author} {\bibfnamefont {C.}~\bibnamefont {Maier}},
  \bibinfo {author} {\bibfnamefont {R.~V.}\ \bibnamefont {Bijnen}}, \bibinfo
  {author} {\bibfnamefont {T.}~\bibnamefont {Brydges}}, \ and\ \bibinfo
  {author} {\bibfnamefont {M.~K.}\ \bibnamefont {Joshi}},\ }\bibfield  {title}
  {\enquote {\bibinfo {title} {{Self-verifying variational quantum simulation
  of lattice models}},}\ }\href {\doibase 10.1038/s41586-019-1177-4} {\bibfield
   {journal} {\bibinfo  {journal} {Nature}\ }\textbf {\bibinfo {volume}
  {569}},\ \bibinfo {pages} {355} (\bibinfo {year} {2019})}\BibitemShut
  {NoStop}%
\bibitem [{\citenamefont {Nielsen}\ and\ \citenamefont
  {Chuang}(2010)}]{Nielsen2010}%
  \BibitemOpen
  \bibfield  {author} {\bibinfo {author} {\bibfnamefont {M.~A.}\ \bibnamefont
  {Nielsen}}\ and\ \bibinfo {author} {\bibfnamefont {I.~L.}\ \bibnamefont
  {Chuang}},\ }\href@noop {} {\emph {\bibinfo {title} {{Quantum Computation and
  Quantum Information}}}},\ \bibinfo {edition} {10th}\ ed.\ (\bibinfo
  {publisher} {Cambridge University Press},\ \bibinfo {address} {Cambridge},\
  \bibinfo {year} {2010})\BibitemShut {NoStop}%
\bibitem [{\citenamefont {Mooij}\ \emph {et~al.}(1999)\citenamefont {Mooij},
  \citenamefont {Orlando}, \citenamefont {Levitov},\ and\ \citenamefont
  {Tian}}]{Mooij1999}%
  \BibitemOpen
  \bibfield  {author} {\bibinfo {author} {\bibfnamefont {J.~E.}\ \bibnamefont
  {Mooij}}, \bibinfo {author} {\bibfnamefont {T.~P.}\ \bibnamefont {Orlando}},
  \bibinfo {author} {\bibfnamefont {L.}~\bibnamefont {Levitov}}, \ and\
  \bibinfo {author} {\bibfnamefont {L.}~\bibnamefont {Tian}},\ }\bibfield
  {title} {\enquote {\bibinfo {title} {{Josephson Persistent-Current Qubit}},}\
  }\href@noop {} {\bibfield  {journal} {\bibinfo  {journal} {Science}\ }\textbf
  {\bibinfo {volume} {285}},\ \bibinfo {pages} {1036--1040} (\bibinfo {year}
  {1999})}\BibitemShut {NoStop}%
\bibitem [{\citenamefont {Yan}\ \emph {et~al.}(2016)\citenamefont {Yan},
  \citenamefont {Gustavsson}, \citenamefont {Kamal}, \citenamefont {Birenbaum},
  \citenamefont {Sears}, \citenamefont {Hover}, \citenamefont {Gudmundsen},
  \citenamefont {Rosenberg}, \citenamefont {Samach}, \citenamefont {Weber},
  \citenamefont {Yoder}, \citenamefont {Orlando}, \citenamefont {Clarke},
  \citenamefont {Kerman},\ and\ \citenamefont {Oliver}}]{Yan2016}%
  \BibitemOpen
  \bibfield  {author} {\bibinfo {author} {\bibfnamefont {F.}~\bibnamefont
  {Yan}}, \bibinfo {author} {\bibfnamefont {S.}~\bibnamefont {Gustavsson}},
  \bibinfo {author} {\bibfnamefont {A.}~\bibnamefont {Kamal}}, \bibinfo
  {author} {\bibfnamefont {J.}~\bibnamefont {Birenbaum}}, \bibinfo {author}
  {\bibfnamefont {A.~P.}\ \bibnamefont {Sears}}, \bibinfo {author}
  {\bibfnamefont {D.}~\bibnamefont {Hover}}, \bibinfo {author} {\bibfnamefont
  {T.~J.}\ \bibnamefont {Gudmundsen}}, \bibinfo {author} {\bibfnamefont
  {D.}~\bibnamefont {Rosenberg}}, \bibinfo {author} {\bibfnamefont
  {G.}~\bibnamefont {Samach}}, \bibinfo {author} {\bibfnamefont
  {S.}~\bibnamefont {Weber}}, \bibinfo {author} {\bibfnamefont {J.~L.}\
  \bibnamefont {Yoder}}, \bibinfo {author} {\bibfnamefont {T.~P.}\ \bibnamefont
  {Orlando}}, \bibinfo {author} {\bibfnamefont {J.}~\bibnamefont {Clarke}},
  \bibinfo {author} {\bibfnamefont {A.~J.}\ \bibnamefont {Kerman}}, \ and\
  \bibinfo {author} {\bibfnamefont {W.~D.}\ \bibnamefont {Oliver}},\ }\bibfield
   {title} {\enquote {\bibinfo {title} {{The flux qubit revisited to enhance
  coherence and reproducibility}},}\ }\href {\doibase 10.1038/ncomms12964}
  {\bibfield  {journal} {\bibinfo  {journal} {Nature Communications}\ }\textbf
  {\bibinfo {volume} {7}},\ \bibinfo {pages} {12964} (\bibinfo {year}
  {2016})}\BibitemShut {NoStop}%
\end{thebibliography}%

\end{document}